\documentclass[]{rsos}

\usepackage{lipsum}           
\usepackage{xargs}           
\usepackage[pdftex,dvipsnames]{xcolor} 
\usepackage[colorinlistoftodos,prependcaption,textsize=small]{todonotes}
\newcommandx{\piotr}[2][1=]{\todo[#1]{PB: #2}}
\newcommandx{\matteo}[2][1=]{\todo[linecolor=red,backgroundcolor=red!25,bordercolor=red,#1]{MM: #2}}
\newcommandx{\ania}[2][1=]{\todo[linecolor=blue,backgroundcolor=blue!25,bordercolor=blue,#1]{ACh: #2}}
\newcommandx{\giancarlo}[2][1=]{\todo[linecolor=blue,backgroundcolor=green!25,bordercolor=blue,#1]{GR: #2}}
\newcommandx{\comment}[2][1=]{\todo[linecolor=Plum,backgroundcolor=Plum!25,bordercolor=Plum,#1]{#2}}

\usepackage{graphicx}
\usepackage{amssymb}			
\usepackage{amsmath}			
\usepackage{amsfonts}
\usepackage{algorithm}
\usepackage{algorithmic}
\usepackage{authblk}
\usepackage{subcaption}
\usepackage{epstopdf}

\usepackage{hyperref}
\hypersetup{
    unicode=false,          
    pdftoolbar=true,        
    pdfmenubar=true,        
    pdffitwindow=false,     
    pdftitle={Blank},    
    pdfauthor={Blank},     
    pdfsubject={Blank},   
    pdfcreator={Blank},   
    pdfproducer={Blank},  
    pdfkeywords={Blank}, 
    pdfnewwindow=true,      
    colorlinks=false,       
    linkcolor=red,          
    citecolor=green,        
    filecolor=magenta,      
    urlcolor=cyan           
}

\begin{document}

\newtheorem{definition}{Definition}
\newtheorem{proposition}{Proposition}  
\newtheorem{property}{Property}
\newtheorem{example}{Example}

\title{Quantifying layer similarity in multiplex networks:\\a systematic study}

\author{
Piotr Br\'odka$^{1}$, Anna Chmiel$^{2}$, Matteo Magnani$^{3}$ and Giancarlo Ragozini$^{4}$}

\address{$^{1}$Department of Computational Intelligence, Wroclaw University of Science and Technology, Poland\\
$^{2}$Faculty of Physics, Warsaw University of Technology, Poland\\
$^{3}$Department of Information Technology, Uppsala University, Sweden\\
$^{4}$Department of Political Science, University of Naples Federico II, Italy}

\subject{Computer science}

\subject{Computer science, computer modelling and simulation, network science, graph theory}


\corres{Piotr Br\'odka\\
\email{piotr.brodka@pwr.edu.pl}}

\begin{abstract}
Computing layer similarities is an important way of characterizing multiplex networks because various static properties and dynamic processes depend on the relationships between layers. We provide a taxonomy and experimental evaluation of approaches to compare layers in multiplex networks. Our taxonomy includes, systematizes and extends existing approaches, and is complemented by a set of practical guidelines on how to apply them.
\end{abstract}


\begin{fmtext}
\section{Introduction}

Multiplex networks provide a simple yet expressive way to model a wide range of physical and social systems as sets of entities connected by multiple types of relationships, that in this paper we also call \emph{layers} following the terminology in \cite{Kivela2014}.
For example, a transport network can be modelled as a set of locations, such as cities or streets, connected by different types of public transport like airplanes, trains, and buses.
Several studies have investigated the connection between layer similarity and other properties of the network. For example, we know from previous research that the relationships between layers have an impact on dynamic processes such as behaviour and information diffusion \cite{Salehi2015survey}.

\end{fmtext}

\maketitle

Being able to measure relationships between layers is also essential to validate models aimed at explaining the formation of empirical multilayer networks \cite{MagnaniSBP2013b,Nicosia2013}.
While the problem of comparing different networks has been thoroughly investigated in the literature \cite{bazzoli1999taxonomy,harland2001taxonomy,Faust2006ComparingStructure,pathan2007taxonomy,bassett2008hierarchical,vanWijk2010ComparingTheory,camarinha2012taxonomy,onnela2012taxonomies,Smith2016EmpiricalSize}, the problem of quantifying layer similarity where the same nodes can be present in multiple layers -- which characterizes multiplex networks -- has not been studied in a systematic and comprehensive way so far.

In the literature, we can find a large number of works using layer similarity measures, but most  use them as a tool to study other phenomena such as multiplex network generation \cite{kim13,MagnaniSBP2013b,Nicosia2013}, link prediction \cite{Jalili2017} and spreading processes \cite{Salehi2015survey}. As a result, different works use the same or very similar approaches presented with different names, the relationships between several of these similarity measures have not been explored, and there are no guidelines on how to quantify layer similarity in multiplex networks, e.g., how to choose the appropriate measure given a specific dataset. In addition, various potentially useful layer comparison measures have not been considered yet.

Therefore, in this paper we provide the following contributions: (\textit{i}) a systematic study of approaches and measures to compute the similarity between layers in multiplex networks, based both on a literature study and on a theoretical framing of the problem; (\textit{ii}) a set of measures that have not been used yet to compare layers, complementing those already defined in the literature; (\textit{iii}) an empirical study of the relationships between different measures, compared on several real datasets, and (\textit{iv}) a set of guidelines on how to choose and use these measures.

In Section~\ref{preliminaries} we present the definitions, concepts, and notation used in the paper. In Section~\ref{similarity} we present an organized set of existing and new layer similarity measures. Section~\ref{empirical} provides the results of an empirical study where the main similarity measures are applied to several real datasets from different domains, such as genetic networks, social networks, co-authorship networks, and transport networks. Section~\ref{guidelines} discusses guidelines to be used to select the most appropriate measure. 


\section{Concepts, terminology and notation}\label{preliminaries}

In this section, we define the basic concepts needed to provide a systematic coverage of layer similarity measures. We start with the standard definition of multiplex network, followed by an alternative representation called property matrix allowing us to define similarity functions based on different types of network structures and different ways to look at them.

\begin{figure}[ht]
\centering
  \includegraphics[width=.5\textwidth]{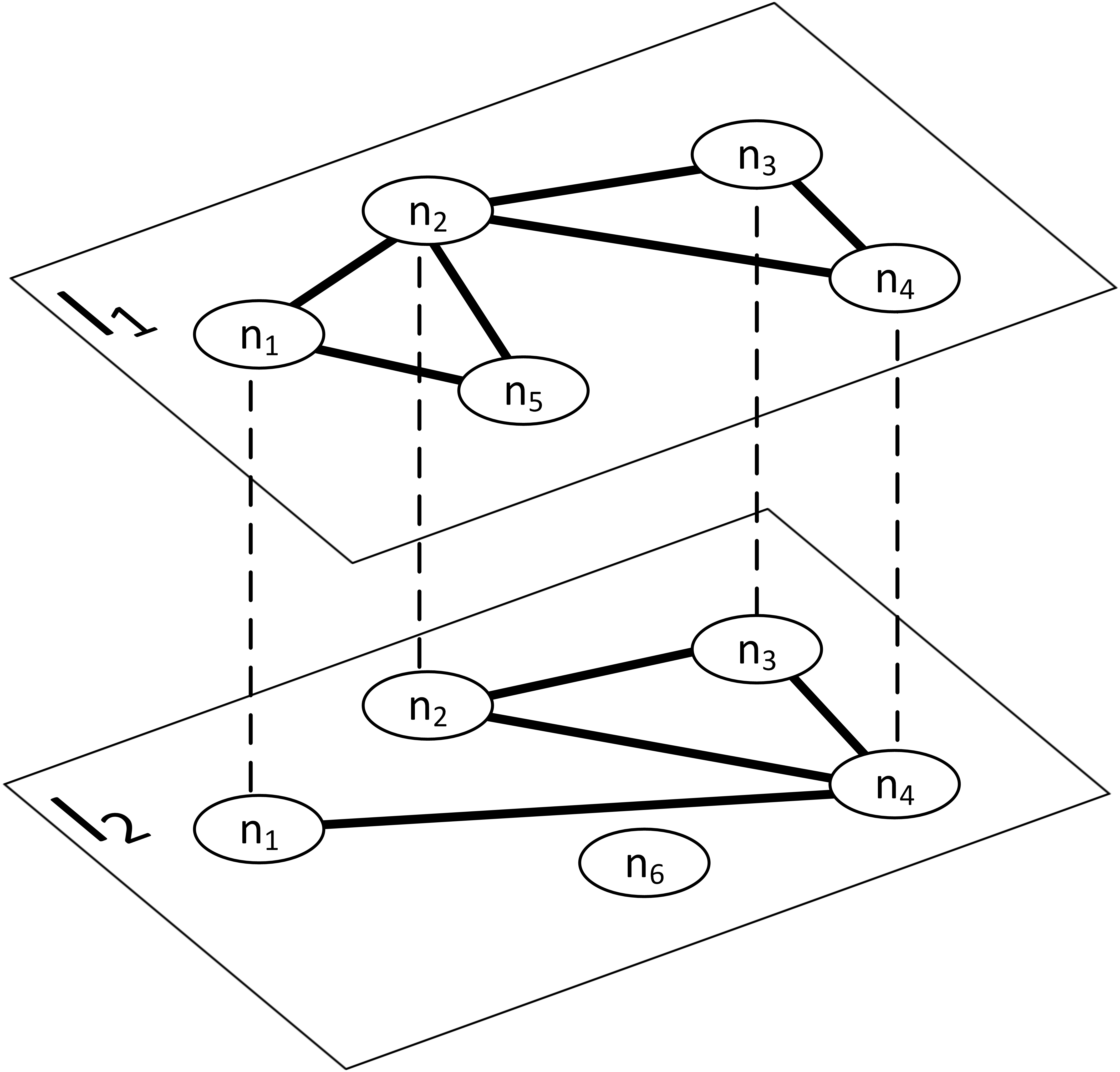}
  \caption{An example of a multiplex network consisting of two layers, six nodes, and ten edges.}
\label{fig:models}
\end{figure}

In this paper we use the following definition of the multiplex network:
\begin{definition}[Multiplex network]
Given a set of nodes $\mathcal{N}$ and a set of layers $\mathcal{L}$, a multiplex network is defined as a quadruple $M = (\mathcal{N},\mathcal{L},V,E)$ where 
$(V,E)$ is a graph,
$V \subseteq \mathcal{N} \times \mathcal{L}$,
and if $(n_1, l_1, n_2, l_2) \in E$ then $l_1=l_2$.
\end{definition}
An example of multiplex network is shown in Figure~\ref{fig:models}, where $\mathcal{L} = \{l_1, l_2\}$, $\mathcal{N} = \{n_1, \dots, n_6\}$, and $(n_1,l_1,n_2,l_1)$ is an example of an edge in $E$. In the literature alternative terminologies are used, and here we adopt the one in \cite{Kivela2014}, according to which we would say that node $n_1$ is present in both layer $l_1$ and layer $l_2$. In the literature some extended multiplex models have also been proposed, allowing multi-dimensional layers \cite{Kivela2014} and one-to-many relationships between nodes in different layers \cite{DBLP:conf/asonam/MagnaniR11}, but we do not consider these extensions here.

Please note that the original definition of multiplex network introduced in the field of Social Network Analysis was more restrictive than the one adopted in this paper. In particular, our definition allows some of the nodes not to be present in some layers. For example, $(n_5,l_2) \notin V$ in Figure~\ref{fig:models}. In some cases, when the term multiplex is used it is assumed that all nodes are present in all layers, and this assumption will often affect the result of layer comparisons. To avoid confusion, in this case, we explicitly talk about a node-aligned multiplex network \cite{Kivela2014}:
\begin{definition}[Node-aligned Multiplex network]
A node-aligned multiplex network is a multiplex network $(\mathcal{N},\mathcal{L},V,E)$ where 
$\forall n \in \mathcal{N}, l \in \mathcal{L} : (n,l) \in V$.
\end{definition}
Notice that treating our working example as a 

\begin{table}
\caption{Terminology and notation used in the paper}
\begin{tabular}{l l p{5cm} }
\hline
Symbol & Name \\
\hline
$\mathcal{N}$ & set of nodes $\{n_1, n_2, ..., n_{|N|}\}$\\
$\mathcal{L}$ & set of layers $\{l_1, l_2, ..., l_{|L|}\}$\\
\hline
$\mathbf{P}$ & property matrix \\
$C$ & set of contexts (e.g., network layers, snapshots, groups)\\
$S$ & set of structures (e.g., nodes, edges, dyads, triangles)\\
$\mathbf{p}_c$ & property vector for context $c \in C$ \\
$\mathbf{p}^s$ & property vector for structure $s \in S$ \\
$p_{s,c}$ & property of $s$ in $c$ (e.g., degree of node $s$ on layer $c$)\\
$p_{C',S'}$ & $p$ restricted to contexts in $C' \subseteq C$ and structures in $S' \subseteq S$\\
\hline
\end{tabular}
\label{tab:notation}
\end{table}

\begin{figure}[ht]
\begin{subfigure}{.45\textwidth}
\begin{tabular}{r | c c c c c c}
  & $n_1$ & $n_2$ & $n_3$ & $n_4$ & $n_5$ & $n_6$\\
  \hline
$n_1$ & 0 & 1 & 0 & 0 & 1 & 0\\
$n_2$ & 1 & 0 & 1 & 1 & 1 & 0\\
$n_3$ & 0 & 1 & 0 & 1 & 0 & 0\\
$n_4$ & 0 & 1 & 1 & 0 & 0 & 0\\
$n_5$ & 1 & 1 & 0 & 0 & 0 & 0\\
$n_6$ & 0 & 0 & 0 & 0 & 0 & 0\\
\end{tabular}
\caption{$\textbf{A}_{l_1}$}
\label{fig:A L1}
\end{subfigure}\hspace{.5cm}
\begin{subfigure}{.45\textwidth}
\begin{tabular}{r | c c c c c c}
  & $n_1$ & $n_2$ & $n_3$ & $n_4$ & $n_5$ & $n_6$\\
  \hline
$n_1$ & 0 & 0 & 0 & 1 & 0 & 0\\
$n_2$ & 0 & 0 & 1 & 1 & 0 & 0\\
$n_3$ & 0 & 1 & 0 & 1 & 0 & 0\\
$n_4$ & 1 & 1 & 1 & 0 & 0 & 0\\
$n_5$ & 0 & 0 & 0 & 0 & 0 & 0\\
$n_6$ & 0 & 0 & 0 & 0 & 0 & 0\\
\end{tabular}
\caption{$\textbf{A}_{l_2}$}
\label{fig:A L2}
\end{subfigure}\\
\caption{Adjacency matrices for both layers of exemplary network (fig.\ref{fig:models})}
\label{fig:adjacency}
\end{figure}

Multiplex networks have usually been represented as a set of adjacency matrices $\textbf{A}_{l}$, one for each layer $l$, where $a_{l}(n_1,n_2) = 1$ if there is an edge between node $n_1$ and node $n_2$~in layer $l$, $a_{l}(n_1,n_2) = 0$ otherwise. 
The adjacency matrices for our working example are shown in Figure~\ref{fig:adjacency}. 

However, this representation is not the most appropriate to define similarity measures, for two main reasons. First, it is incomplete, because it only allows representing node-aligned multiplex networks. An example of why this is important is the case of online social media, where each layer represents a different service (Twitter, Facebook, etc.) and it makes a difference whether a user has no connections on Twitter or does not even have an account there. In our working example, we would lose the information that nodes $n_5$ and $n_6$ are present in different layers.

Second, adjacency matrices present an edge-oriented view over the multiplex network, which might be the reason why most similarity measures in the literature have been limited to edge similarity. If we take a broader look at empirical networks, we can see how other structures can be relevant. As an example, if we look at Figure~\ref{fig:models} we can see that the triangle $\{n_2, n_3, n_4\}$ is present in both layers. Unfortunately, this is not obvious from the adjacency matrices and would require checking several disparate entries making definitions more complicated than needed. 
Therefore, in the following, we use network representation targeted to the specific properties we want to consider when checking the similarity between layers. We call this representation a \emph{property matrix}. 

\begin{definition}[Property matrix]
A \emph{property matrix} $\mathbf{P}$ is a matrix where:
\begin{enumerate}
\item the columns correspond to a set $S$ of network structures (nodes, edges, triangles, \dots),
\item the rows correspond to a set $C$ of contexts where these structures are observed (layers, groups, snapshots, \dots), and
\item $p_{s,c}$ is the value of an observational function mapping each pair structure/context into a number (degree, distance, \dots).
\end{enumerate}
\end{definition}

Since in this paper we focus on layer similarity we will only use layers as contexts, that is, $C = \mathcal{L}$. In summary, each cell $p_{s,c}$ of a property matrix contains the value of the function describing the structure $s$ (for example, a node) on layer $c$, and different observational functions can be used to define different types of similarity. Examples of property matrices for our working example are shown in Figure~\ref{fig:property matrices}.

\begin{figure}[ht]
\begin{subfigure}{.5\textwidth}
\begin{tabular}{r | c c c c c c}
  & $n_1$ & $n_2$ & $n_3$ & $n_4$ & $n_5$ & $n_6$\\
\hline
$l_1$ & 1 & 1 & 1 & 1 & 1 & 0\\
$l_2$ & 1 & 1 & 1 & 1 & 0 & 1 \\
\hline
\end{tabular}
\caption{Nodes, existence}
\label{fig:pm clique1}
\end{subfigure}
\begin{subfigure}{.5\textwidth}
\begin{tabular}{r | c c c c c c}
  & $n_1$ & $n_2$ & $n_3$ & $n_4$ & $n_5$ & $n_6$\\
\hline
$l_1$ & 2 & 4 & 2 & 2 & 2 & NA\\
$l_2$ & 1 & 2 & 2 & 3 & NA & 0 \\
\hline
\end{tabular}
\caption{Nodes, degree}
\label{fig:pm degree}
\end{subfigure}
\begin{subfigure}{.5\textwidth}
\begin{tabular}{r | c c c c c c}
  & $n_1$ & $n_2$ & $n_3$ & $n_4$ & $n_5$ & $n_6$\\
\hline
$l_1$ & 2 & 1/3 & 1 & 1 & 1 & NA\\
$l_2$ & 1 & 1 & 1 & 1/3 & NA & 0 \\
\hline
\end{tabular}
\caption{Nodes, CC}
\label{fig:pm cc}
\end{subfigure}
\begin{subfigure}{.5\textwidth}
\begin{tabular}{r | c c c c}
  & $(n_1,n_2)$ & \dots & $(n_2,n_4)$ & \dots \\
\hline
$l_1$ & 1 & \dots & 1 & \dots \\
$l_2$ & 0 & \dots & 1 & \dots \\
\hline
\end{tabular}
\caption{Dyads, edge existence (clique)}
\label{fig:pm clique2}
\end{subfigure}
\begin{subfigure}{\textwidth}
\centering
\begin{tabular}{r | c c c c c c c c}
  & $n_1,n_2,n_3$ & \dots & $n_1,n_2,n_5$ & \dots & $n_1,n_2,n_6$ & \dots & $n_2,n_3,n_4$ & \dots \\
\hline
$l_1$ & 0 & \dots & 1 & \dots & 0 & \dots & 1 & \dots \\
$l_2$ & 0 & \dots & 0 & \dots & 0 & \dots & 1 & \dots \\
\hline
\end{tabular}
\caption{Triads, triangle existence (clique)}
\label{fig:pm clique3}
\end{subfigure}
\caption{\textit{Property matrices} for our working example in fig.~\ref{fig:models}. Each \textit{property matrix} is defined by a type of structures (nodes, dyads, triads, etc.), the contexts (layers) and an observational function (existence, degree, forming a clique, distance, etc.)}
\label{fig:property matrices}
\end{figure}

Given a structure $s$, we can further summarize its presence in the network by summing over all the values in $\mathbf{p}^s$, computing their standard deviation or performing any other kind of aggregation (sum, avg, median, min, max, etc.). As an example, from a node-degree property matrix (Figure~\ref{fig:pm degree}) we can obtain the total degree of a node in the whole multiplex network (sum) or its so-called \emph{degree deviation} \cite{Berlingerio2013}, which is 0 if a node has the same number of connections on all layers and higher when a node is present in different layers with different degrees, and so on. In summary, property matrices provide a more general and informative representation of multiplex networks than adjacency matrices -- which are still useful when the objective is just to know about the edges in a node-aligned network. Property matrices also allow us to provide simple and general mathematical definitions of different ways to compare layers, which will instantiate into several existing and new measures when specific property matrices are used.


\section{Layer similarity functions}\label{similarity}

Given a \emph{property matrix} $\mathbf{P}$ where each row represents a layer, we can compare two layers in three main ways. The first is to summarize each row using an aggregation function $f$ and compare $f(\mathbf{p}_{l_1})$ to $f(\mathbf{p}_{l_2})$. For example, if the property matrix contains node degrees we can compare the layers' average degrees $\textrm{mean}(\mathbf{p}_{l_1})$ and $\textrm{mean}(\mathbf{p}_{l_2})$.
Comparing the distribution of values in $\mathbf{p}_{l_1}$ and $\mathbf{p}_{l_2}$ is the second way to compare layers. As an example, we can compare degree distributions on different layers and find that both fit well a power law distribution with the same exponent.
The third way is to compare $p_{s,{l_1}}$ with $p_{s,{l_2}}$ for all $s$. As an example, we can compute degree correlation to check whether nodes with a high (resp., low) degree on one layer tend to have a high (resp., low) degree also on the other layer.

\subsection{Comparing aggregations of layer property vectors}
\label{comp_summary}
This first class of comparison methods is based on comparing $f(\mathbf{p}_{l_1})$ to $f(\mathbf{p}_{l_2})$ using various functions ($f$) aggregating each layer into a single value. Typical choices are basic statistical summary functions such as mean, max, sum, skewness and kurtosis, combinations of the simple statistics, such as the coefficient of variation (the ratio between the standard deviation and the mean), the Jarque-Bera statistics (a combination of skewness and kurtosis), or the Shannon entropy \cite{shannon2002mathematical} of the distribution. These methods are summarized in Table \ref{tab:comp_summary}. 

Then, given $f(\mathbf{p}_{l_1})$ and $f(\mathbf{p}_{l_2})$ we can compare them, and in our experiments we have used their relative difference, i.e. $2\cdot(| f(\mathbf{p}_{l_1}) - f(\mathbf{p}_{l_2})|) / (|f(\mathbf{p}_{l_1})| + |f(\mathbf{p}_{l_2})|)$.


Notice that depending on the property matrix these measures correspond to various existing network summaries. 
For example, the mean function may return the average degree (when applied to property matrices about node degrees, or the global clustering coefficient also known as transitivity index (for node clustering coefficients), or the average path length for property matrices about dyads and geodesic distances.
--- which in the field of chemistry coincides with the Wiener index 
\cite{wiener1947structural}.

\begin{table}[!h]
\caption{Summary of common aggregation functions for \emph{property matrices}}
\label{tab:comp_summary}
\centering
\begin{tabular}{l l}
\hline
Name & Function \\
\hline
$mean(\mathbf{p}_{l})$ & $\frac{\sum_s  p_{s,l}}{card(\mathbf{p}_{l})}$ \\
$sd(\mathbf{p}_{l})$& $\sqrt{\frac{\sum_s \left( p_{s,l} - mean(\mathbf{p}_{l}) \right)^2}{card(\mathbf{p}_{l})}}$ \\
$skew(\mathbf{p}_{l})$ & $ \frac{\sum_s \left(_{s,l} - mean(\mathbf{p}_l) \right)^3}{card(\mathbf{p}_l) sd(\mathbf{p}_l)^3  }$ \\
$kurt(\mathbf{p}_l)$ & $ \frac{\sum_s \left( p_{s,l} - mean(\mathbf{p}_l) \right)^4}{card(\mathbf{p}_l) sd(\mathbf{p}_l)^4  }$ \\
$entropy(\mathbf{p}_l)$ & $\sum_{k=1}^k fr_{k,l} \log fr_{k,l}$ \\
$CV(\mathbf{p}_l)$ & $\frac{sd(\mathbf{p}_l)}{mean(\mathbf{p}_l)}$  \\
$Jarque-Bera(\mathbf{p}_l)$ & $\frac{card(\mathbf{p}_l)}{6} \left( skew(\mathbf{p}_l)^2 + \frac{\left( kurt(\mathbf{p}_l)-3 \right)^2}{4} \right) $  \\
\hline
\multicolumn{2}{p{10cm}}{$fr_{k,l}$ is the relative frequency of the $k$-th value of the property vector $\mathbf{p}_l$ in a generic layer $l$}
\end{tabular}
\end{table}

\subsection{Comparing distributions of layer property vectors}\label{comp_distribution}


While using a single value to compare layers can provide some useful knowledge about the multiplex network, for example by highlighting the presence of denser or more clustered layers than others, looking at the whole distribution of values in the property matrix can reveal other types of relationships among layers. From a statistical point of view, some ways are open to pursuing this task. The first one consists in comparing the moments of two distributions. For example, it is possible to compare the first four moments, even if by theoretical point of view this is not completely sufficient.
Another possible approach consists in comparing the distributions directly. In this case, we have to apply to each property vector a function $\textrm{fr}(\mathbf{p}_l)$ that derives the relative frequency distribution. In case of discrete distributions, such as the degree distribution, given a property vector $\mathbf{p}_l$ we derive the disjoint values $p_{k,l}$, $k=1, \ldots, K$, and we associate to each value the relative frequency $fr_{k,l}$.

In case of continuous distribution, or in case of very large networks in which also the discrete distributions take a wide range of values, the function $\textrm{fr}(\mathbf{p}_l)$ derives histograms. We first divide the range of values of the property vector into $K$ equal interval, or bins, $[b_{(k-1)}, b_{k}]$, with $b_{0}$ being the minimum value in the property matrix and $b_{K,l}$ being the maximum value in the property matrix\footnote{If we only compare two rows, we can also choose the minimum and maximum values in those rows.}. 
Then we associate the relative frequency $fr_{k}$ to each interval. Note that the bins of all histograms for all layers must be the same. Then we have to compare only the relative frequency distributions. This procedure is very fast and efficient also for very large networks.

Given the frequencies or histograms, in order to compare two layers we can use the distance between observed distributions based on distance between histograms, namely, the dissimilarity index ($ID$), the Kullback-Leibler divergence $D_{KL}$ \cite{kullback1951information}, the Jensen-Shannon divergence $D_{JS}$ or the Jeffrey divergence $D_{J}$, as defined in Table \ref{tab:comp_distributions} \cite{Crooks2008}. In the following, we do not consider the Jeffrey divergence, as the Jensen-Shannon divergence is its smother version.
Note that this kind of comparison can be made both for node-aligned and for not node-aligned multiplexes.

\begin{table}[!h]

\caption{Main methods to compare distributions across layers}
\label{tab:comp_distributions}
\centering
\begin{tabular}{l l l}
\hline
Name & Notation & Function \\
\hline
Dissimilarity index & $ID(\mathbf{p}_{l_1},\mathbf{p}_{l_2})$ & $\frac{1}{2} \sum_{k=1}^K | fr_{k,{l_1}} - fr_{k,{l_2}} |$ \\
Kullback-Leibler & $D_{KL}(\mathbf{p}_{l_1},\mathbf{p}_{l_2})$& $\sum_{k=1}^K fr_{k,{l_1}} \log \frac{fr_{k,{l_1}}}{fr_{k,{l_2}}}$ \\
Jensen-Shannon & $D_{JS}(\mathbf{p}_{l_1},\mathbf{p}_{l_2})$ & $ \frac{1}{2}(\sum_{k=1}^K fr_{k,{l_1}} \log \frac{fr_{k,{l_1}}}{{\hat fr}_k} + fr_{k,{l_2}} \log \frac{fr_{k,{l_2}}}{{\hat fr}_k})$  \\
Jeffrey & $D_{J}(\mathbf{p}_{l_1},\mathbf{p}_{l_2})$ & $\sum_{k=1}^K fr_{k,{l_1}} \log \frac{fr_{k,{l_1}}}{fr_{k,{l_2}}} + \sum_{k=1}^K fr_{k,{l_2}} \log \frac{fr_{k,{l_2}}}{fr_{k,{l_1}}}$ \\
\hline\\
\multicolumn{2}{l}{${\hat fr}_k = \frac{fr_{k,{l_1}} +fr_{k,{l_2}}}{2}$}
\end{tabular}
\end{table}

\subsection{Comparing individual structures}

The main feature of multiplex networks is that the same structure can be present or not, and have different characteristics, on each layer. For example, a node can be present in one layer and not in the other, or the same node may have different degrees depending on the layer. Therefore, a peculiar set of measures to compare layers relies on the comparison of the structures of interest, one by one. 

Two main cases are possible. In property matrices indicating the existence of different structures on the different layers, we only have two values, 0 and 1. While represented as numbers, these are in fact just nominal values indicating that the structure is present on the layer. For these binary matrices specific methods can be used, checking the overlapping or more in general, the common existence (or common absence) of structures across layers. For numerical matrices containing generic numbers, e.g., node degrees, other methods are more appropriate, as described in the following two sections.

\subsubsection{Binary properties}

When a structure can be present or not on different layers, a basic way to compute the similarity between layers is to quantify the overlapping of these structures, that is, how often the same structure appears or not on more than one layer. This is typically the case when the observation function defining the property matrix checks the existence of the structure.

Measures of overlapping have been defined and redefined many times during the last few years in different papers, but most definitions can be generalized using property matrices as:
\begin{equation}\label{overlapping_eq}
C \mathbf{p}'_{l_1}\cdot \mathbf{p}_{l_2},
\end{equation}
where $C$ is some normalization function. Most (but not all) measures in the literature compare edges across layers, this being the result of the traditional edge-based definitions of multiplex networks such as adjacency matrices. In our definition, the usage of \emph{property matrices} allows us to apply similar comparisons to various other properties.

Consider two binary property vectors $\mathbf{p}_{l_1}$ and $\mathbf{p}_{l_2}$. Following \cite{Batagelj1995} let us denote with:
\begin{itemize}
\item[-] $a= \mathbf{p}'_{l_1}\cdot \mathbf{p}_{l_2}$ the number of properties that $l_1$ and $l_2$ share;
\item[-] $b= \mathbf{p}'_{l_1}\cdot (\mathbf{1}-\mathbf{p}_{l_2})$ the number of properties that $l_1$ has and $l_2$ lacks;
\item[-] $c= (\mathbf{1}-\mathbf{p}'_{l_1}) \cdot \mathbf{p}_{l_2}$ the number of properties that $l_1$ lacks and $l_2$ has;
\item[-] $d= (\mathbf{1}-\mathbf{p}_{l_1})' \cdot (\mathbf{1}-\mathbf{p}_{l_2})$ the number of properties that both $l_1$ and $l_2$ lacks;
\item[-] $m=a+b+c+d=length(\mathbf{p}_{l_1})=length(\mathbf{p}_{l_2})$
\end{itemize}
Then, the binary similarity functions can be summarized as follow in Table \ref{tab:comp_binary}.
\begin{table}[!h]
\centering
\caption{Similarity functions for binary property matrices. Column $C$ indicates the normalization function in Eq.~\ref{overlapping_eq}. For the two functions also considering the non-existence of structures on both layers, we only provide the standard definition not based on the product of property vectors}
\label{tab:comp_binary}
\centering
\begin{tabular}{l l l}
\hline
Name &  Normalization function $C$ & Standard notation\\
\hline
Russel-Rao &  $\frac{1}{length(\mathbf{p}_{l_1})}$ & $\frac{a}{m}$\\
Jaccard &  $\frac{1}{length(\mathbf{p}_{l_1}) - (\mathbf{1}-\mathbf{p}_{l_1})' \cdot (\mathbf{1}-\mathbf{p}_{l_2})}$ & $\frac{a}{m-d}$ \\
Coverage & $\frac{1}{length(\mathbf{p}_{l_1})}$  &\\
Kulczy\'nski &  $\frac{1}{ 2 }(\frac{1}{\parallel \mathbf{p}_{l_1}\parallel_1}+\frac{1}{\parallel \mathbf{p}_{l_2}\parallel_1})$& $\frac{a}{2}(\frac{1}{a+b} + \frac{1}{a+c} )$\\
\hline
Simple matching coefficient (SMC) & NA & $\frac{a+d}{m}$ \\
Hamann & NA & $\frac{a+d-(b+c)}{m}$ \\
\hline
\end{tabular}
\end{table}


\subsubsection{Numerical properties}

Depending on the reason why we are computing the similarity between layers, we can use different approaches. As each layer is represented as a vector in a property matrix, one way is to compute vectorial distances such as Euclidean distance or cosine similarity.
Another popular way to compare numerical layer property vectors is to compute correlations. An example of this is the so-called inter-layer correlation measure, which is just the Pearson coefficient computed on two node degree property vectors \cite{Berlingerio2012,Nicosia2015}. It is interesting to notice that in the literature correlations across layers have been almost always computed on node degrees, and in \cite{battiston2014structural} also on clustering coefficients.
However, correlations can be in fact be computed on any \textit{property matrix}.

When computing correlations in generalized multiplex networks a choice must be made on how to handle actors not present in all layers. The choice we adopted in our experiments was to discard pairs where at least one of the two values was missing, which is a typical option in statistical software packages.


\begin{table}[!h]

\caption{Similarity functions for numerical property matrices. The function $\rho(\cdot)$ provides the ranks of the values in the property vectors}
\label{tab:comp_numerical}
\centering
\begin{tabular}{l l}
\hline
Name & Function \\
\hline
Cosine Similarity & $\frac{\mathbf{p}'_{l_1} \cdot \mathbf{p}_{l_2}}{\parallel \mathbf{p}_{l_1} \parallel \cdot \parallel \mathbf{p}_{l_2} \parallel }$ \\
Person Correlation Coefficient & $\frac{[\mathbf{p}_{l_1}-mean(\mathbf{p}_{l_1})]' \cdot [\mathbf{p}_{l_2}-mean(\mathbf{p}_{l_2})]}{\parallel [\mathbf{p}_{l_1}-mean(\mathbf{p}_{l_1})] \parallel \cdot \parallel [\mathbf{p}_{l_2}-mean(\mathbf{p}_{l_2})] \parallel }$  \\
Spearman Correlation Coefficient & $\frac{[\rho (\mathbf{p}_{l_1})-mean(\rho (\mathbf{p}_{l_1}))]' \cdot [\rho(\mathbf{p}_{l_2})-mean(\rho(\mathbf{p}_{l_2}))]}{\parallel [\rho(\mathbf{p}_{l_1})-mean(\rho(\mathbf{p}_{l_1}))] \parallel \cdot \parallel [\rho(\mathbf{p}_{l_2})-mean(\rho(\mathbf{p}_{l_2}))] \parallel }$  \\
\hline\\
\end{tabular}
\end{table}


\section{Empirical comparison of measures}\label{empirical}

The experiments have been performed using the multinet library\footnote{\url{https://cran.r-project.org/package=multinet}} and twenty-three multilayer networks\footnote{\url{http://deim.urv.cat/~manlio.dedomenico/data.php}}. 

\begin{table}[!ht]
\centering
\caption{Twenty-three multilayer networks used during experiments}
\label{tab:networks} 
\begin{tabular}{l|p{4.6cm}|p{2,5cm}|l|l}
\textbf{ID} & \textbf{Network} & \textbf{Description} & \textbf{\# of layers} & \textbf{Ref.}\\
\hline
1 & Bos Linnaeus & Genetic & 4 & \cite{stark2006biogrid} \\
2 & Candida Albicans & Genetic & 7 & \cite{stark2006biogrid} \\
3 & Celegans & Genetic, & 6 & \cite{stark2006biogrid}\\
4 & Danio Rerio & Genetic & 5 & \cite{stark2006biogrid} \\
5 & Gallus Gallus & Genetic & 6 & \cite{stark2006biogrid} \\
6 & Hepatitus C & Genetic & 3 & \cite{stark2006biogrid} \\
7 & Human Herpes Virus & Genetic & 4 & \cite{stark2006biogrid} \\
8 & Human HIV Virus & Genetic & 5 & \cite{stark2006biogrid} \\
9 & Oryctolagus & Genetic & 3 & \cite{stark2006biogrid} \\
10 & Plasmodium Falciparum & Genetic & 3 & \cite{stark2006biogrid} \\
11 & Rattus Norvegicus & Genetic & 6 & \cite{stark2006biogrid} \\
12 & Xenopus Laevis & Genetic & 5 & \cite{stark2006biogrid} \\
\hline
13 & Ckm Physicians Innovation & Social & 3 & \cite{coleman1957diffusion} \\ 
14 & Cs Aarhus & Social & 5 & \cite{rossi2015towards}\\ 
15 & Florentine Families & Social & 2 & \cite{padgett1993robust} \\
16 & Kapferer Tailor Shop & Social & 4 & \cite{kapferer1972strategy} \\
17 & Krackhardt High Tech & Social & 3 & \cite{krackhardt1987cognitive} \\
18 & Lazega Law Firm &  Social & 3 & \cite{snijders2006new} \\
19 & Vickers Chan $7^{th}$graders & Social & 2 & \cite{vickers1981representing} \\
\hline
20 & Arxiv Network Science & Co-authorship & 13 & \cite{de2015identifying} \\ 
21 & Pierre Auger & Co-authorship & 16 & \cite{de2015identifying} \\ 
\hline
22 & EU Air Transportation & Transport & 37 & \cite{cardillo2013emergence} \\
23 & London Transport & Transport & 3 & \cite{de2014navigability} \\ 
\hline
\end{tabular}
\end{table}

When we indicate "node-aligned", all the nodes have been replicated into all layers. Otherwise, a node is added to a layer only if the node has at least one edge on that layer -- in which case we talk of generalized multiplex networks. The input format of the multinet library allows the distinction between nodes without connections and missing nodes, as in our working example, but none of the datasets we have used explicitly make this distinction.

\begin{figure}
     \centering
    
    \begin{subfigure}[t]{0.49\textwidth}
        \raisebox{-\height}{\includegraphics[width=\textwidth]{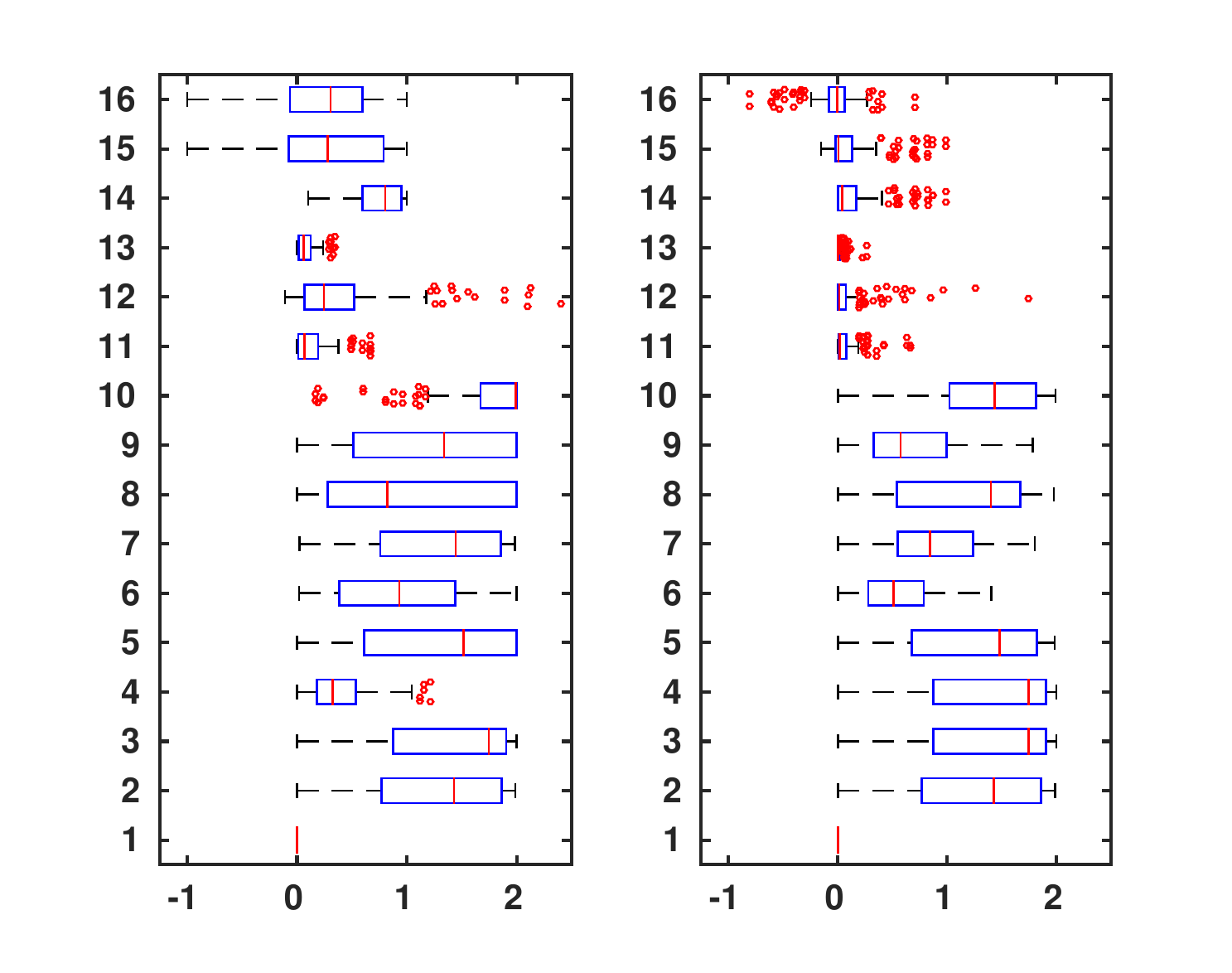}}
        \caption{Genetic networks}
		\label{fig:genetic_box d}
    \end{subfigure}
    \hfill
    \begin{subfigure}[t]{0.50\textwidth}
        \raisebox{-\height}{\includegraphics[width=\textwidth]{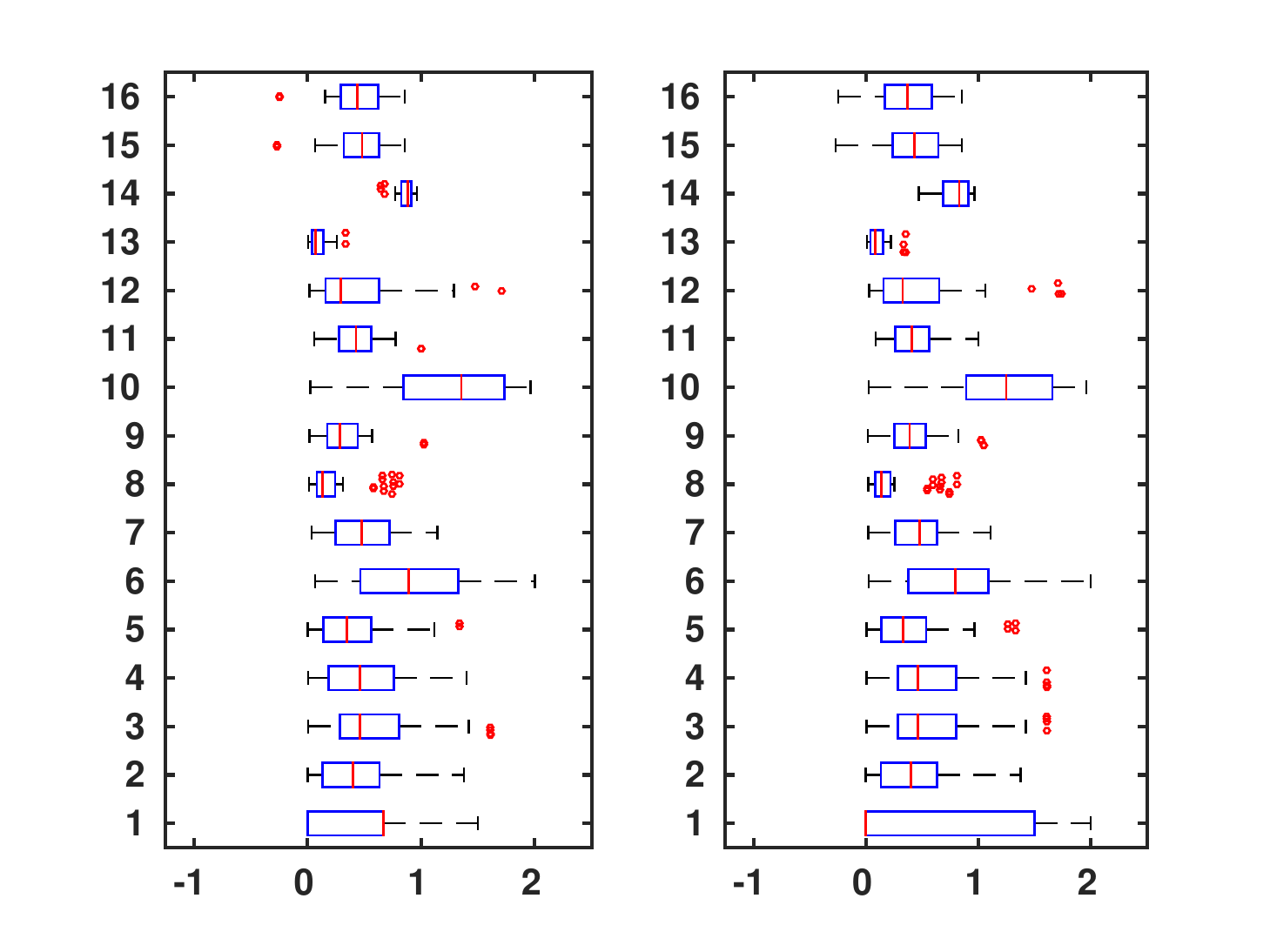}}
        \caption{Social networks}
		\label{fig:social_box d}
    \end{subfigure}
     \begin{subfigure}[t]{0.49\textwidth}
        \raisebox{-\height}{\includegraphics[width=\textwidth]{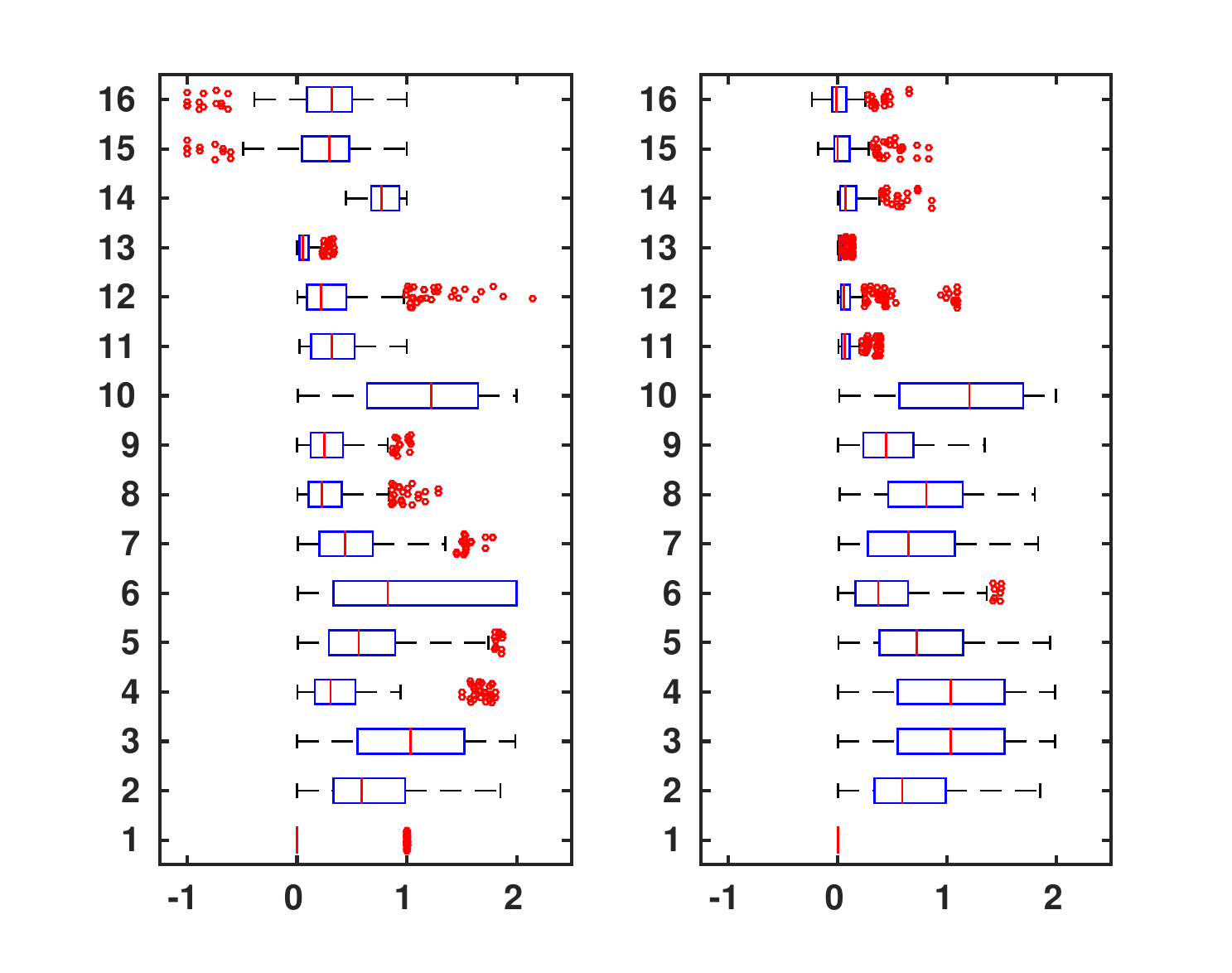}}
        \caption{Co-authorship networks}
		\label{fig:colaboration_box d}
    \end{subfigure}
    \hfill
    \begin{subfigure}[t]{0.49\textwidth}
        \raisebox{-\height}{\includegraphics[width=\textwidth]{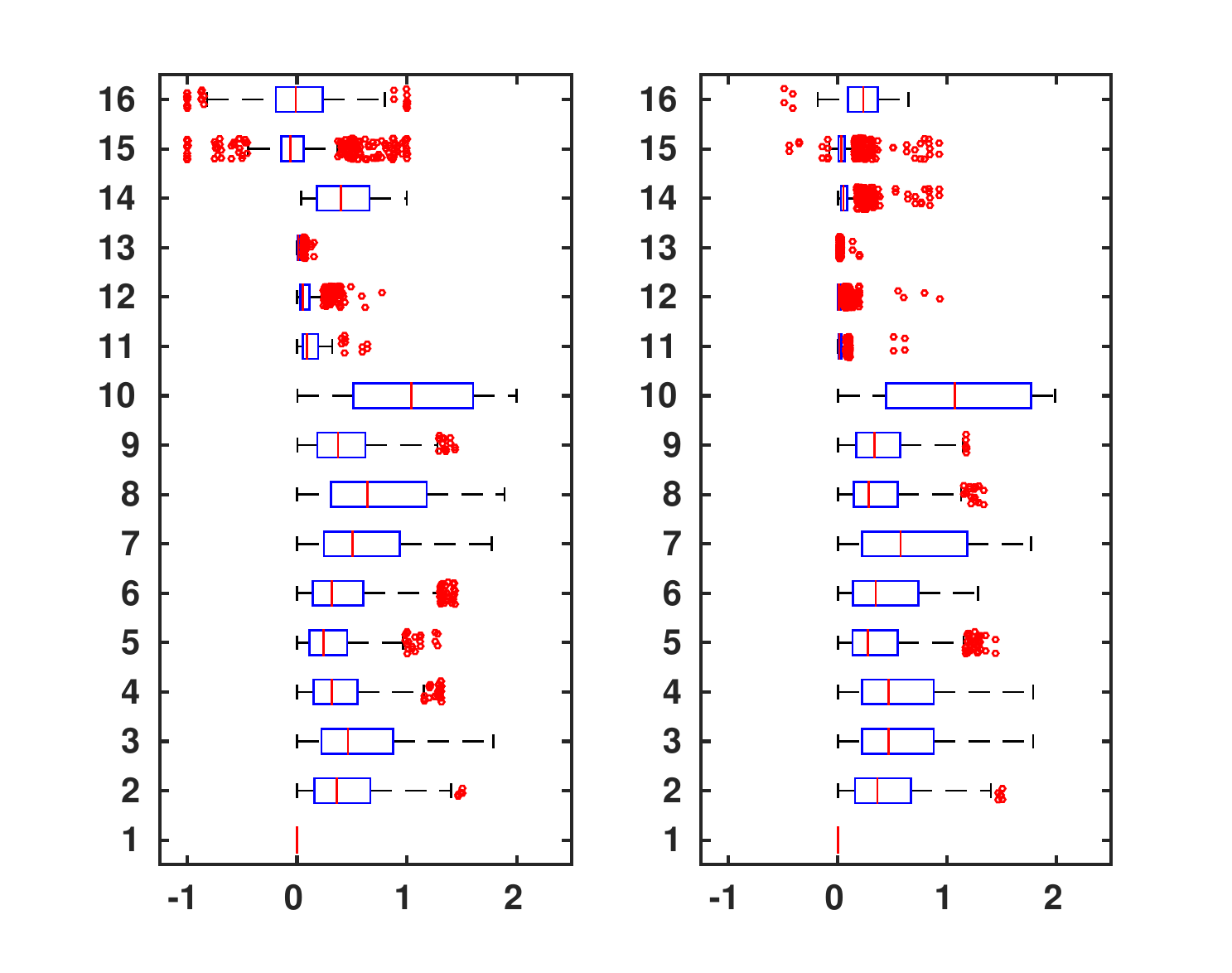}}
        \caption{Transport networks}
		\label{fig:transportation_box d}
    \end{subfigure}
   
    \caption{Boxplots for degree based measures (1-16). For each network type on the left is generalized multiplex network and on the right the node-aligned multiplex network. The outliers have been scattered.}
    \label{Results degree}
\end{figure}

\begin{figure}
     \centering
    \begin{subfigure}[t]{0.49\textwidth}
        \raisebox{-\height}{\includegraphics[width=\textwidth]{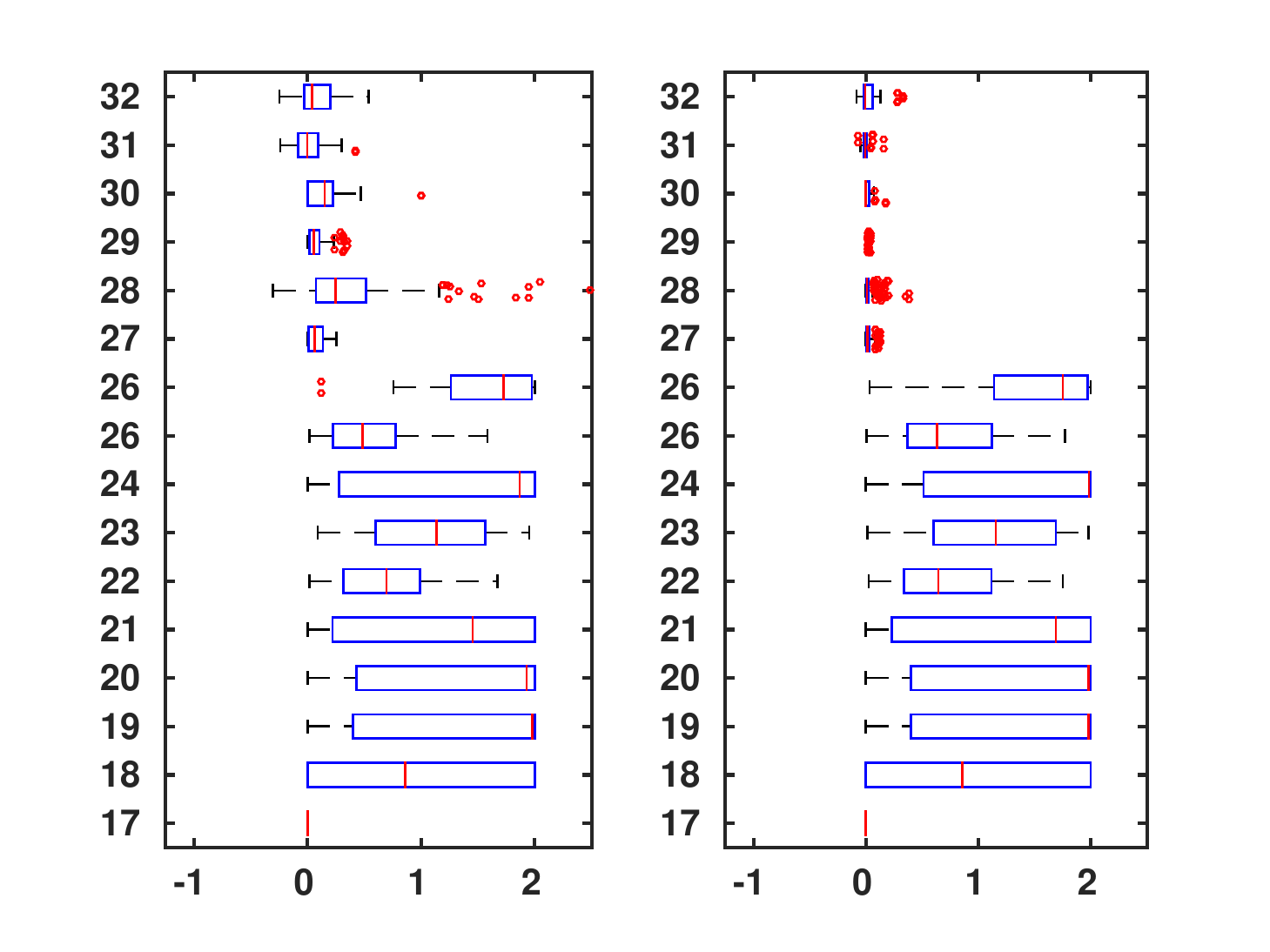}}
        \caption{Genetic networks}
		\label{fig:genetic_box cc}
    \end{subfigure}
    \hfill
    \begin{subfigure}[t]{0.49\textwidth}
        \raisebox{-\height}{\includegraphics[width=\textwidth]{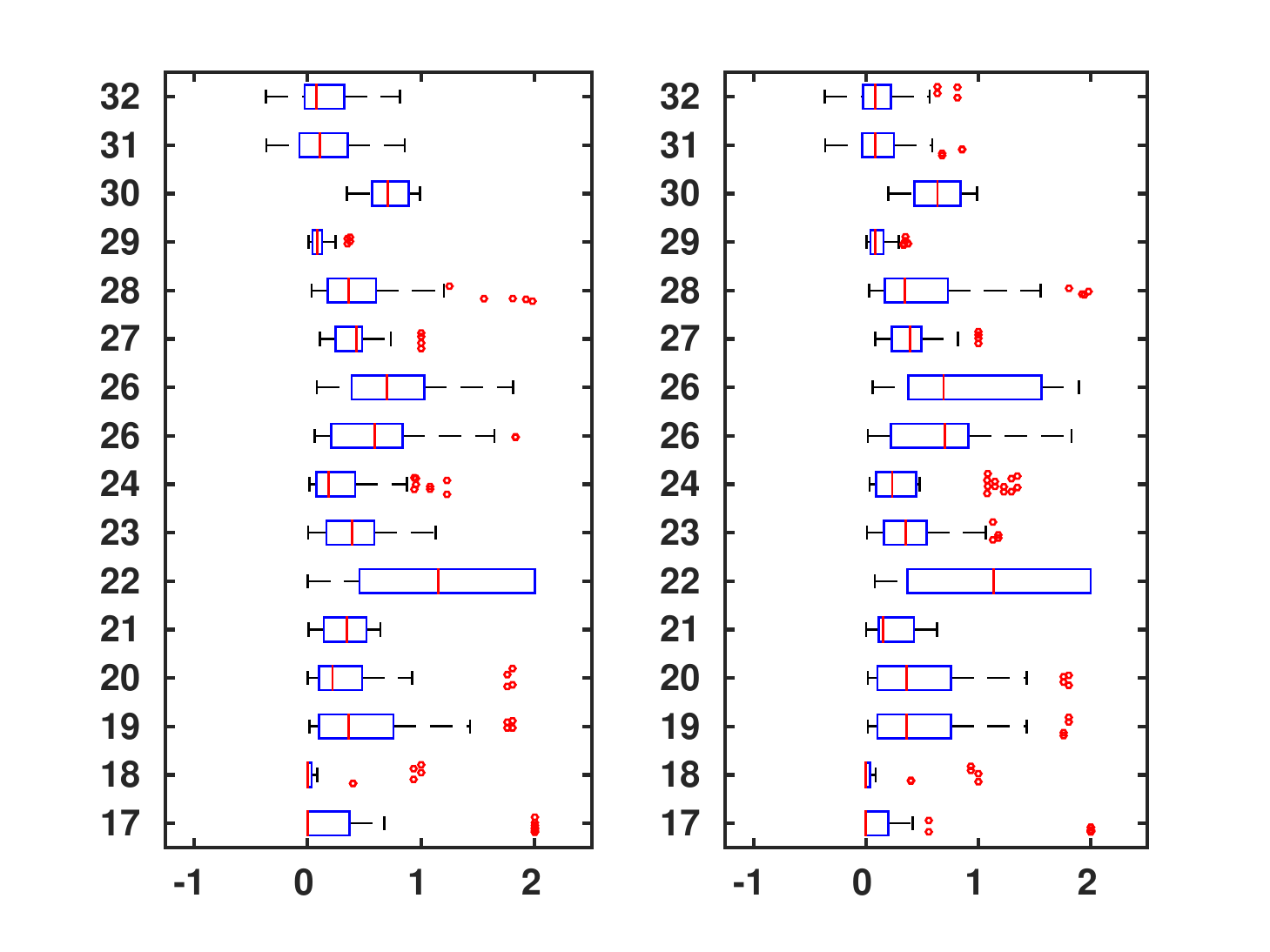}}
        \caption{Social networks}
		\label{fig:social_box cc}
    \end{subfigure}
    \begin{subfigure}[t]{0.49\textwidth}
        \raisebox{-\height}{\includegraphics[width=\textwidth]{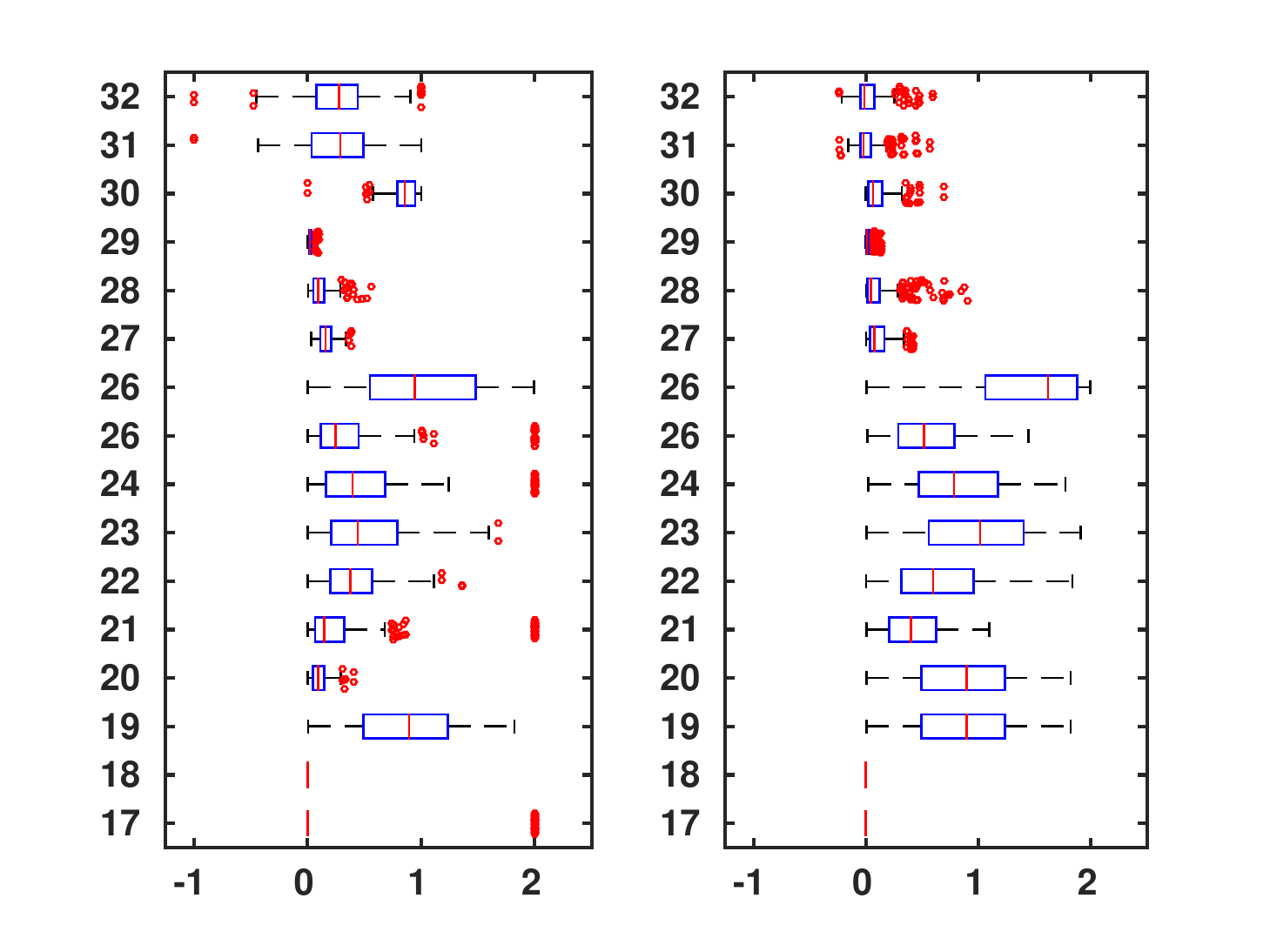}}
        \caption{Co-authorship network}
		\label{fig:colaboration_box cc}
    \end{subfigure}
    \hfill
    \begin{subfigure}[t]{0.49\textwidth}
        \raisebox{-\height}{\includegraphics[width=\textwidth]{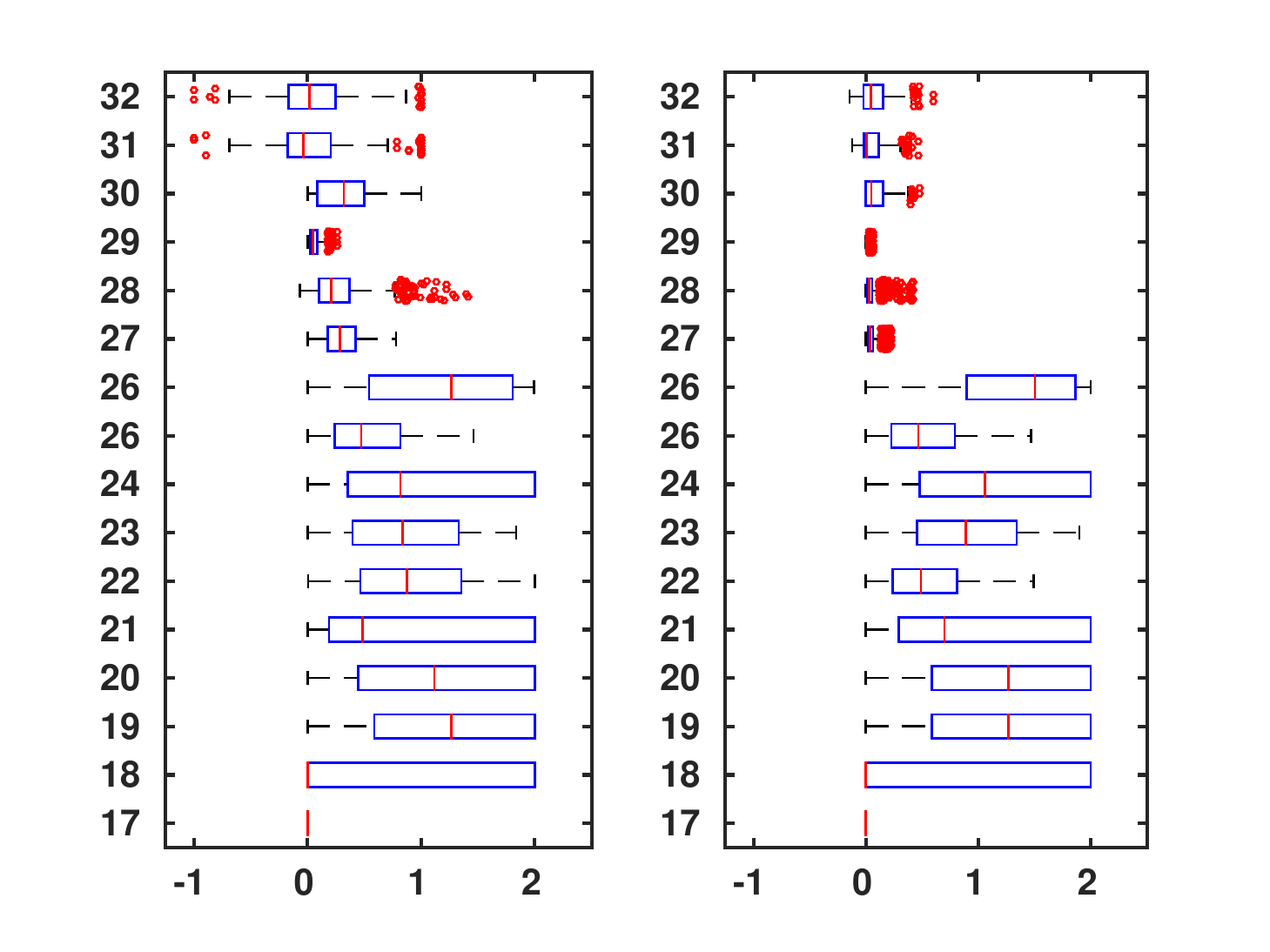}}
        \caption{Transport networks}
		\label{fig:transportation_box cc} 
    \end{subfigure}
   
    \caption{Boxplots for clustering coefficient based measures (17-32). For each network type on the left is generalized multiplex network and on the right the node-aligned multiplex network. The outliers have been scattered.}
    \label{Results CC}
\end{figure}

\begin{figure}
     \centering
    \begin{subfigure}[t]{0.49\textwidth}
        \raisebox{-\height}{\includegraphics[width=\textwidth]{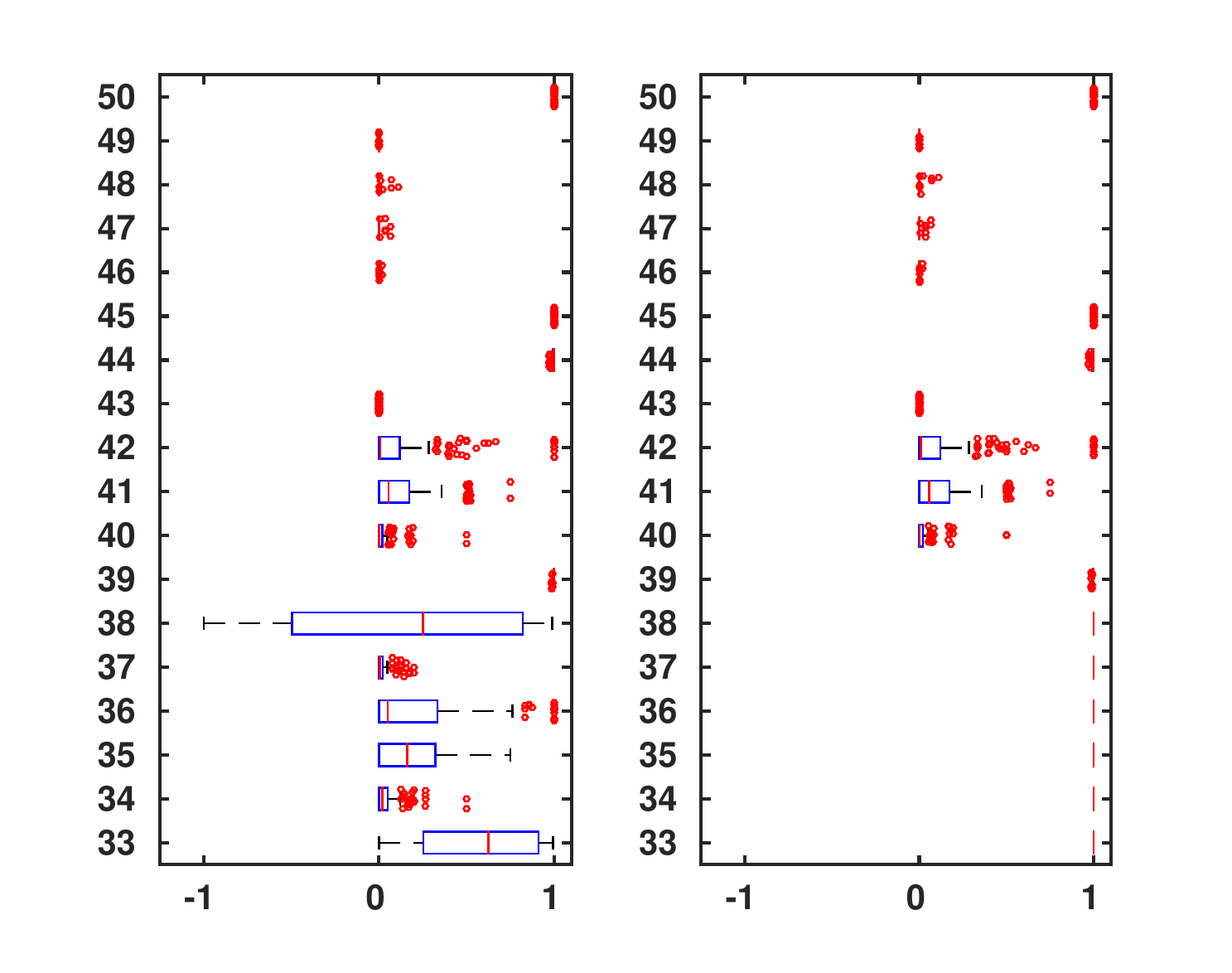}}
        \caption{Genetic networks}
		\label{fig:genetic_box o}
    \end{subfigure}
    \hfill
    \begin{subfigure}[t]{0.49\textwidth}
        \raisebox{-\height}{\includegraphics[width=\textwidth]{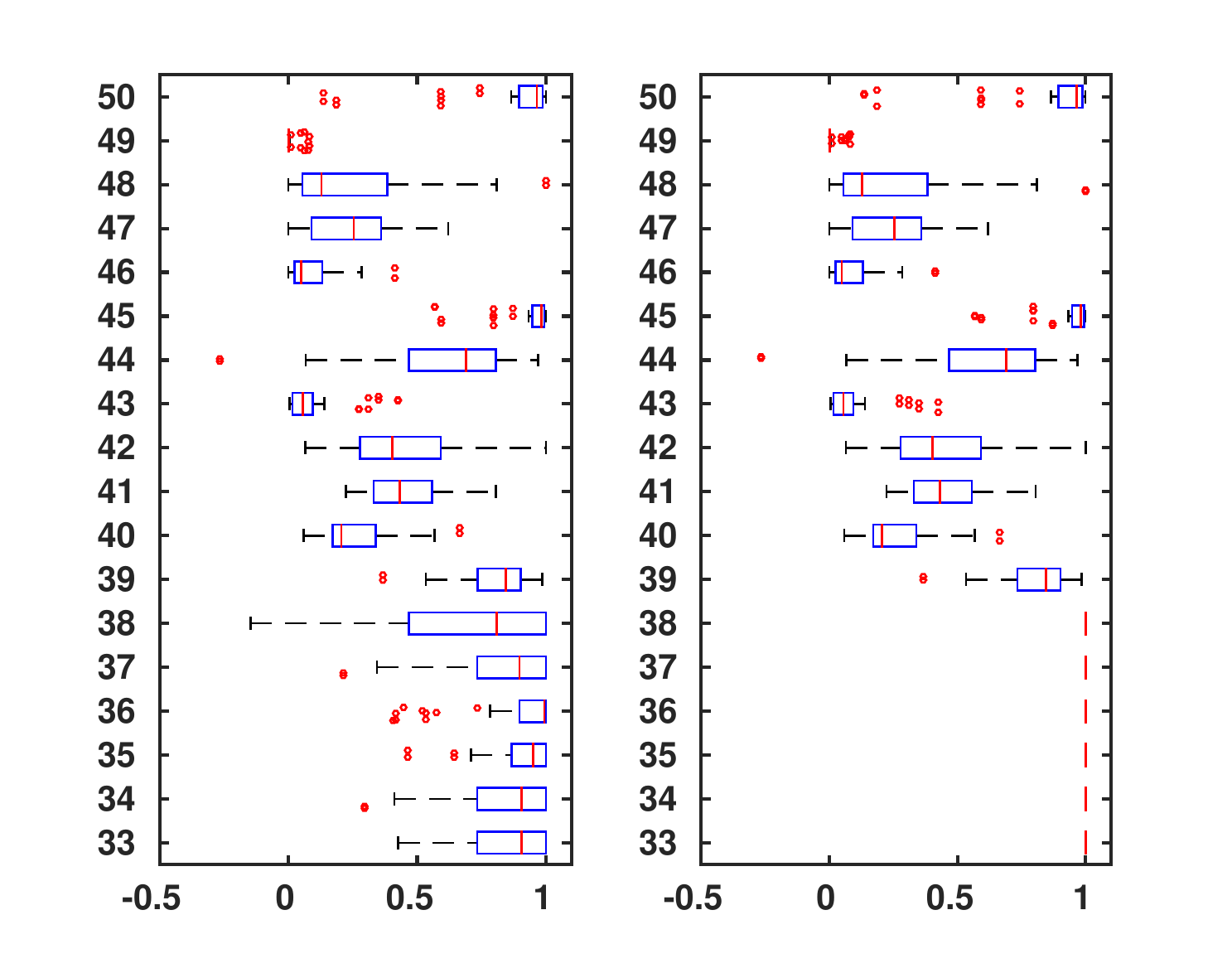}}
        \caption{Social networks}
		\label{fig:social_box o}
    \end{subfigure}
    \begin{subfigure}[t]{0.49\textwidth}
        \raisebox{-\height}{\includegraphics[width=\textwidth]{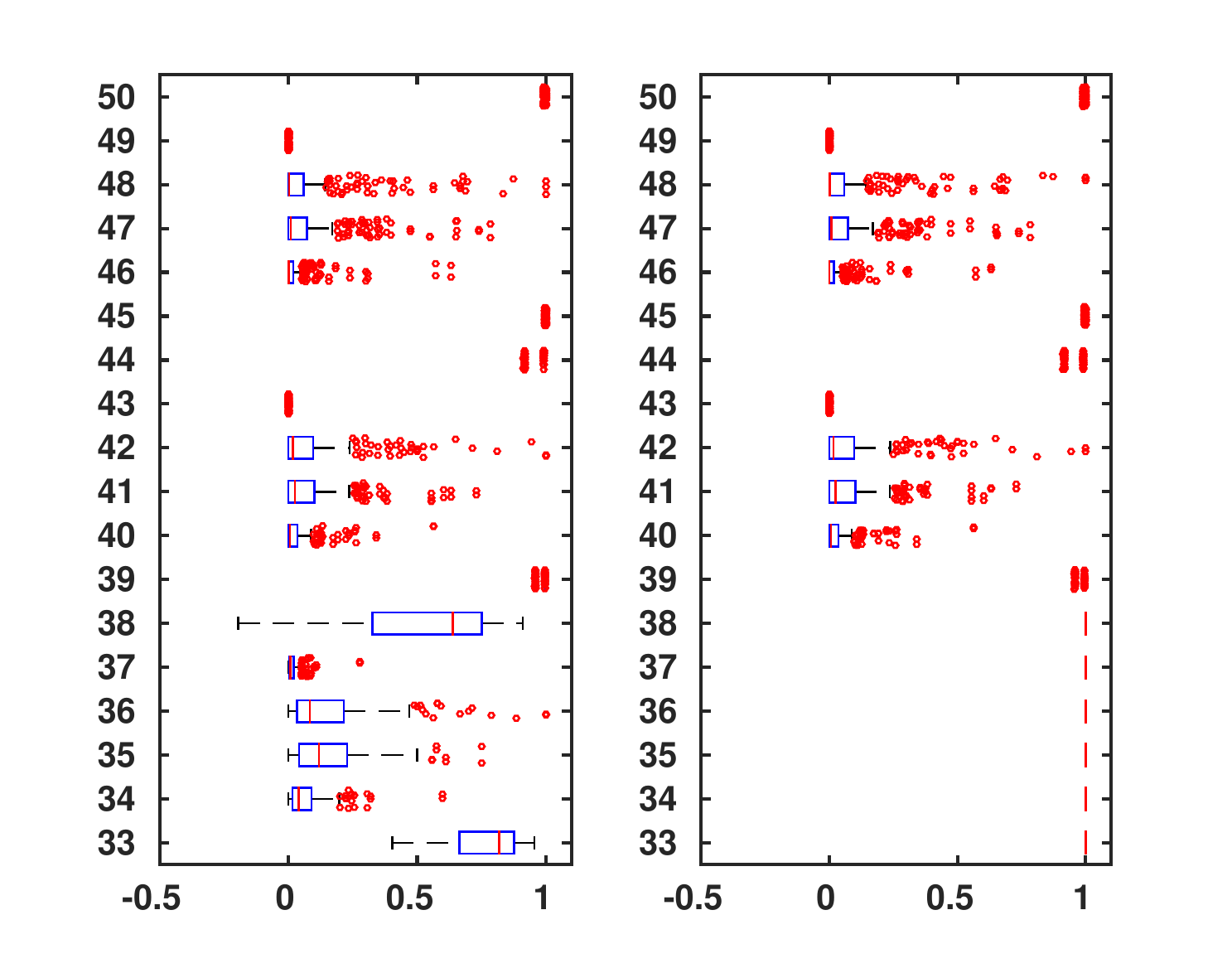}}
        \caption{Co-authorship network}
		\label{fig:colaboration_box o}
    \end{subfigure}
    \hfill
    \begin{subfigure}[t]{0.49\textwidth}
        \raisebox{-\height}{\includegraphics[width=\textwidth]{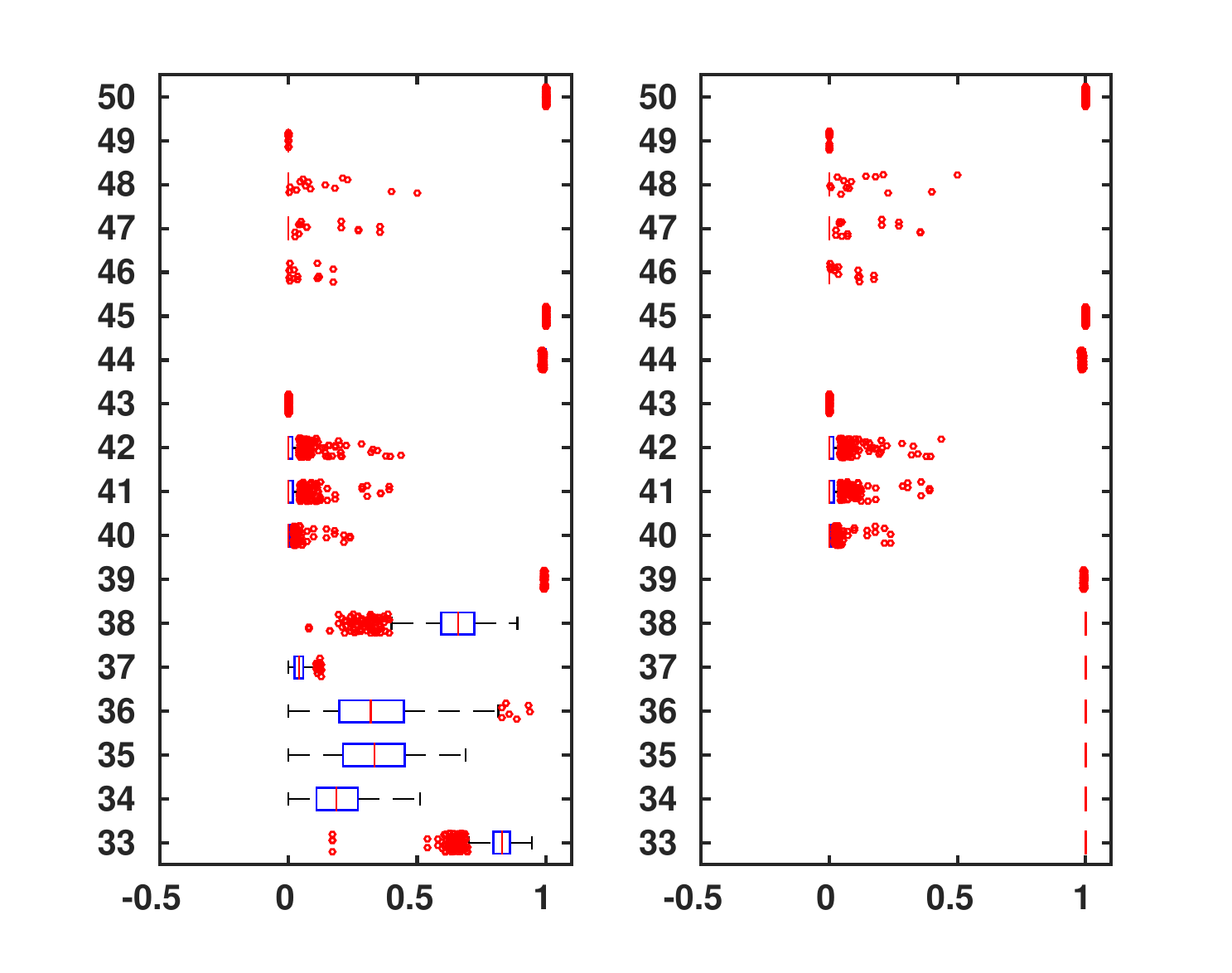}}
        \caption{Transport networks}
		\label{fig:transportation_box o}
    \end{subfigure}
   
    \caption{Boxplots for node-, edge- and triangle-based measures (33-50). For each network type on the left is generalized multiplex network and on the right the node-aligned multiplex network. The outliers have been scattered.}    \label{Results overlapping}
\end{figure}

\begin{figure}
     \centering
        \raisebox{-\height}{\includegraphics[width=\textwidth]{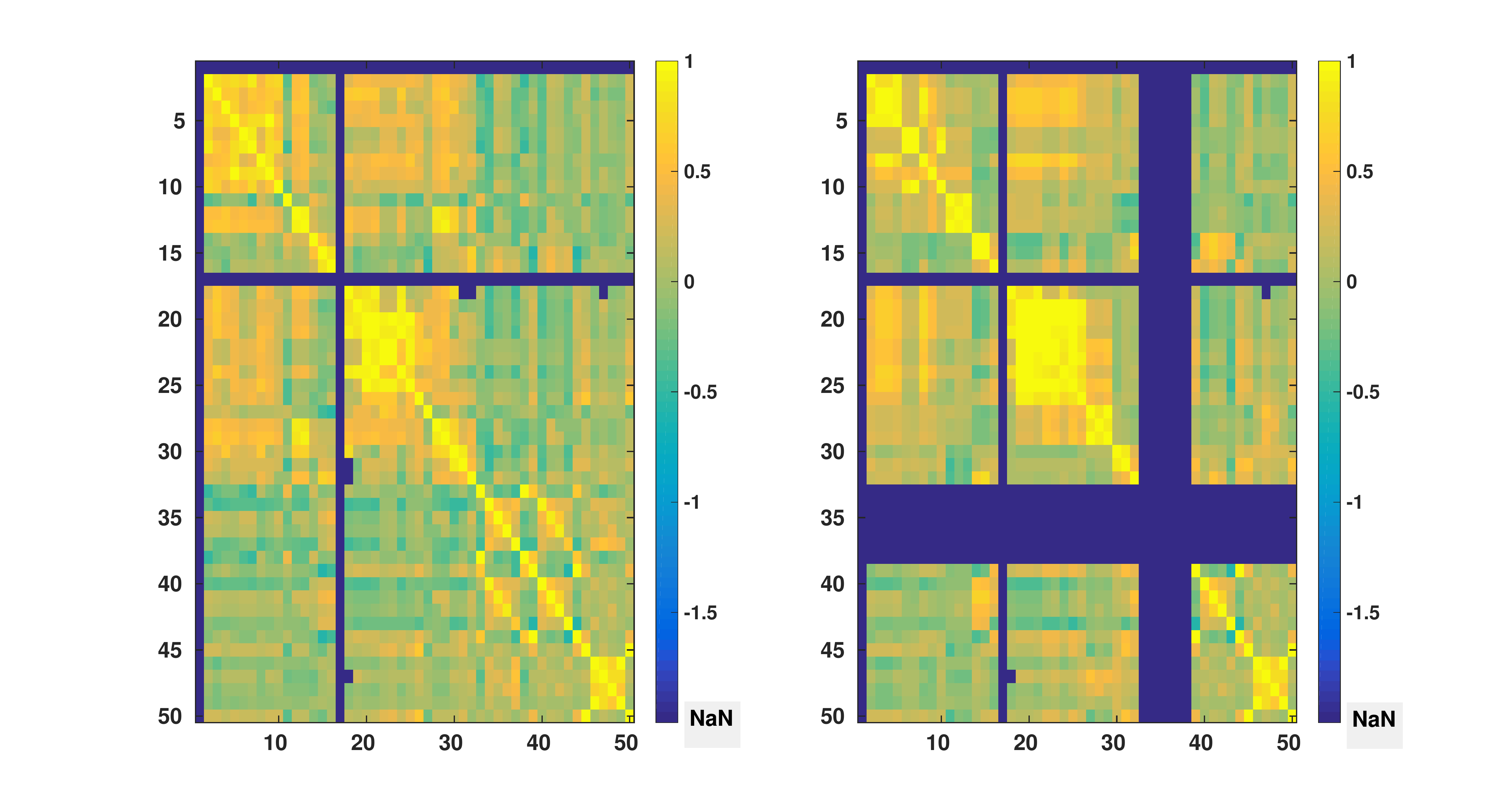}}
        \caption{Correlation between all fifty measures for Genetic networks. On the left is generalized multiplex network and on the right the node-aligned multiplex network}
		\label{fig:genetic_box corr}
    \end{figure}

\begin{figure}
     \centering
        \raisebox{-\height}{\includegraphics[width=\textwidth]{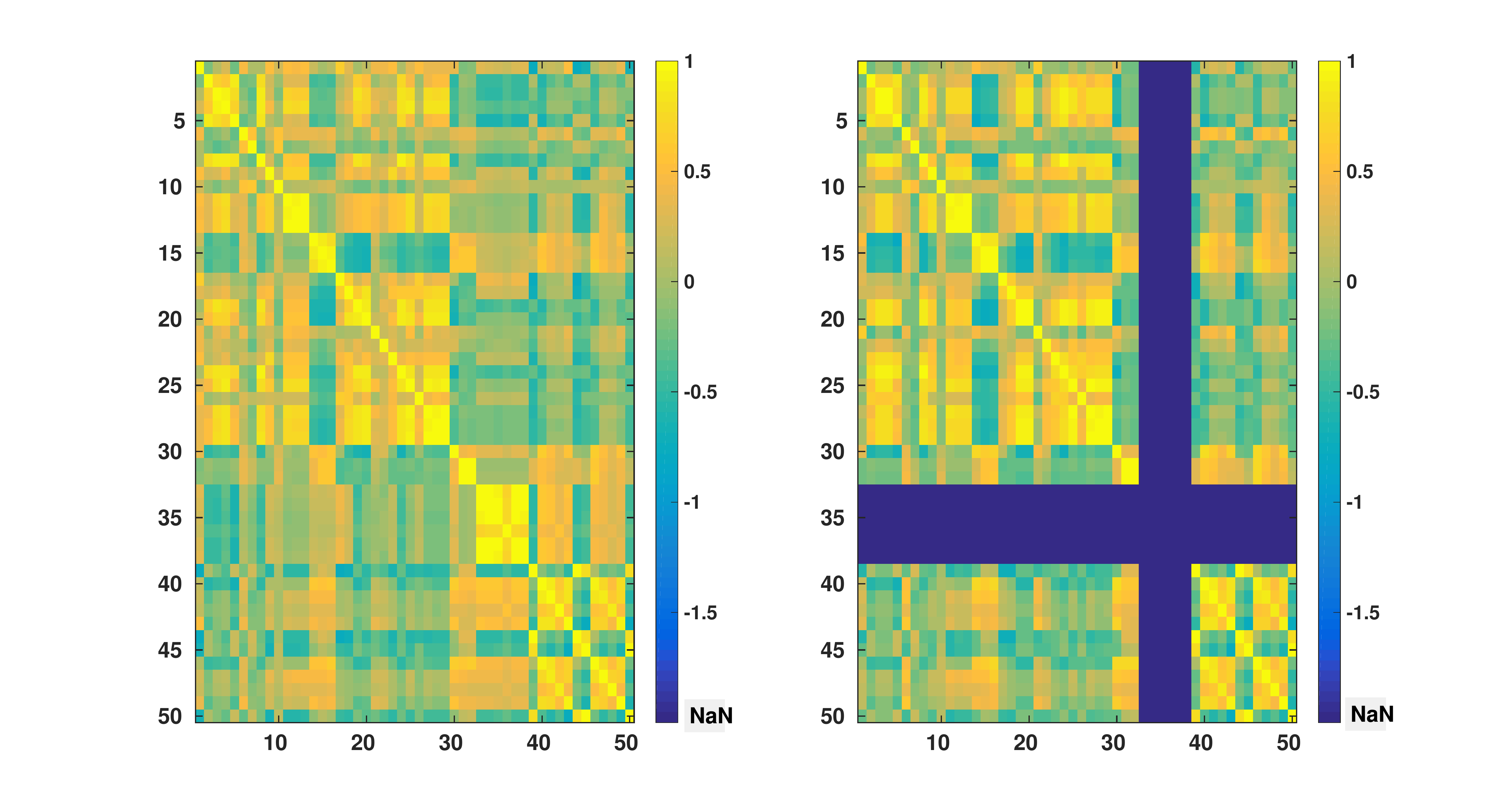}}
        \caption{Correlation between all fifty measures for Social networks. On the left is generalized multiplex network and on the right the node-aligned multiplex network}
		\label{fig:social_box corr}
    \end{figure}

\begin{figure}
     \centering
        \raisebox{-\height}{\includegraphics[width=\textwidth]{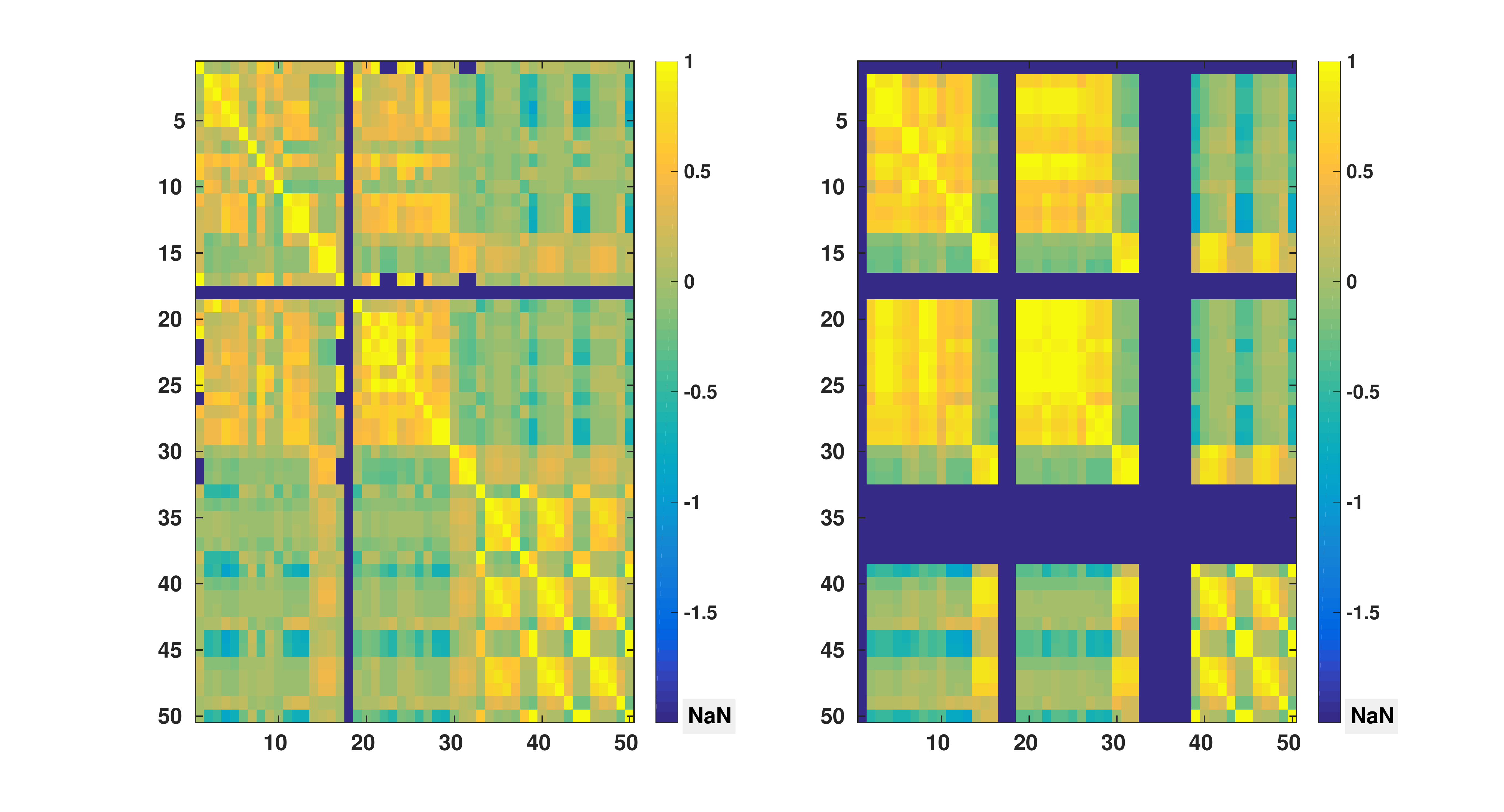}}
        \caption{Correlation between all fifty measures for Co-authorship networks. On the left is generalized multiplex network and on the right the node-aligned multiplex network}
		\label{fig:collaboration_box corr}
    \end{figure}

\begin{figure}
     \centering
        \raisebox{-\height}{\includegraphics[width=\textwidth]{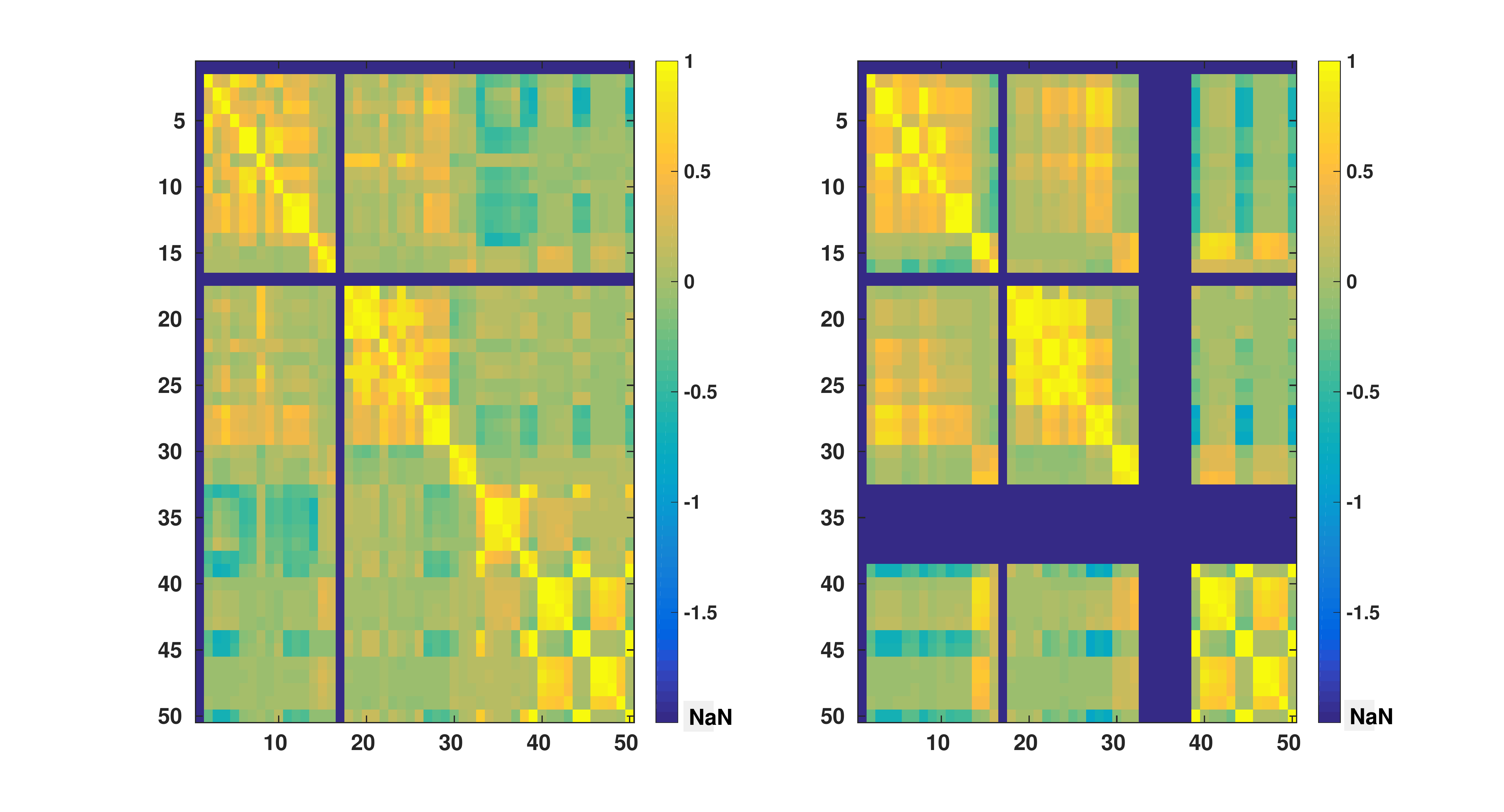}}
        \caption{Correlation between all fifty measures for Transport networks. On the left is generalized multiplex network and on the right the node-aligned multiplex network}
		\label{fig:transport_box corr}
    \end{figure}

In the experiments, we have computed the similarity between all pairs of layers in each dataset and grouped these results by network type (Table~\ref{tab:networks}). Figures \ref{Results degree}, \ref{Results CC} and  \ref{Results overlapping} show the properties of distribution of values produced by each measure. Figures \ref{fig:genetic_box corr}, \ref{fig:social_box corr}, \ref{fig:transport_box corr} and \ref{fig:collaboration_box corr} show the Pearson correlation between values obtained by different measures, where a value of 1 (yellow in the colour figures) indicates that two measures are equivalent (up to some constant rescaling). In addition to the results presented in these figures, we have also performed a manual qualitative analysis of the results, to verify our interpretation of the patterns emerging in the plots.

In the following sections, we highlight some of the results, grouped into three main areas.

\begin{table}
\caption{Fifty measures evaluated during experiments}
\label{tab:mapping}
\begin{tabular}{|l|l|l|l|l|l|}
1& min degree&17& min CC&33& SMC node\\
2& max degree&18& max CC&34& Jaccard node\\
3& sum degree&19& sum CC&35& Kulczy\'nski node\\
4& mean degree&20& mean CC&36& coverage node\\
5& standard deviation degree&21& standard deviation CC&37& Russel-Rao node\\
6& skewness degree&22& skewness CC&38& Hamann node\\
7& kurtosis degree&23& kurtosis CC&39& SMC edge\\
8& entropy degree&24& entropy CC&40& Jaccard edge\\
9& CV degree&25& CV CC&41& Kulczy\'nski edge\\
10& Jarque-Bera degree&26& Jarque-Bera CC&42& coverage edge\\
11& Dissimilarity index degree&27& Dissimilarity index CC&43& Russel-Rao edge\\
12& KL divergence degree&28& KL divergence CC&44& Hamann edge\\
13& JS divergence degree&29& JS divergence CC&45& SMC triangle\\
14& Cosine distance degree&30& Cosine distance CC&46& Jaccard triangle\\
15& Pearson correlation degree&31& Pearson correlation CC&47& Kulczy\'nski triangle\\
16& Spearman correlation degree&32& Spearman correlation CC&48& coverage triangle\\
  &  & &  &49& Russel-Rao triangle\\
&&&&50& Hamann triangle\\
\end{tabular}
\end{table}


\subsection{Overlapping-based measures}

Overlapping-based measures have been used multiple times in the literature, mainly applied to edges. In Figure \ref{Results overlapping} we can observe their behaviours on the various datasets used in our experiments.

Measures based on Simple Matching, Russel-Rao and Hamann degenerate whenever the property vectors become large (that is, $m$ is large) and sparse (that is, $d$ is close to $m$). In these cases, Russel-Rao tends to 0 while Hamann and SMC tend to 1, as we can see in the plots. However, with node-existence property matrices, these degeneration conditions are often not verified, so these measures can still capture different levels of similarity.

When applied to generalized multiplex networks, node overlapping shows significant differences between different types of networks. For example, in Figure \ref{fig:social_box o} we can see that social networks tend to have a high node overlapping (average close to one for measures 34-36), while for example, co-authorship networks show values closer to 0, indicating a significant difference between people working in different disciplines (Figure \ref{fig:colaboration_box o}). In practice, we can say that many social networks are naturally node-aligned.

However, in both cases, we can see several outliers, highlighting special relationships between layers and thus showing the usefulness of these measures also to identify special cases. For example for the Arxiv co-authorship network (20 in table \ref{tab:networks}) two layers physics.data-an (Physics Data Analysis, Statistics and Probability) and cs.SI (Computer science Social and Information Networks) are very similar in terms of node overlapping, indicating an interdisciplinary topic which is of interest to both computer scientists and physicists. 
Another example, this time for social networks, comes from the AUCS network (14 in table \ref{tab:networks}). Almost all outliers are related to the two layers facebook and co-author, both having a significantly different number of actors if compared with the other layers in the network, which explains, e.g., low overlapping. 

Higher order structures, that is, dyads and triads in our experiments, also show different behaviours in different types of networks. There are several similar layers in collaboration networks, may be because that these networks are often obtained as projections from bipartite networks, but still, the majority of the pairs of layers are not very similar. For social networks, a high overlapping is observed much more frequently, also because of the high presence of triangles, while transportation and genetic networks show the least overlapping.

\subsection{Correlation-based measures}

Correlation measures (15, 16, 31, 32) prove their usefulness by discriminating between, e.g., social networks, where the degrees are correlated --- that is, (un)popular people are often (un)popular on more than one layer, while for co-authorship networks where layers indicate different disciplines researchers are often popular in one or a few of them. Interestingly, transport networks contain different extremes: airports that are hubs for one airline are often not hubs for others (corresponding to anti-correlations, that is, values towards 1 in the figures) while for the London data the same locations are often hubs for different types of transportation, resulting in positive correlations.

In case of social networks and co-authorship networks, we often find a positive correlation between degree-based measures (1-16) and measures based on the clustering coefficient (17-32). This is not the case for genetic and transport networks.

In many cases, Pearson and Rank correlations show similar results.

\subsection{Effects of node alignment}

The impact of using a node-aligned or generalized multiplex is evident in many experimental results, as expected. Obviously, node-based measures computing the overlapping among nodes in different layers (33-38) become useless if we force all layers to contain all nodes (Figure \ref{Results overlapping}, right-hand-side plots).

At the same time, using node-aligned networks also affects many other measures. As an example, Figure \ref{fig:transportation_box d} shows the presence of anti-correlated layers (measures 15 and 16, left-hand-side, values close to -1), revealing how airports that are hubs for one airline are often not hubs for others. Considering many nodes that would not be present in the layers, and thus having degree 0, makes these anti-correlations less evident (measures 15 and 16, right-hand-side, values now closer to 0). 

For edge- and triangle-based overlapping measures the results are the same in the node-aligned and in the non-aligned networks. This, however, only because we have not made a difference between, e.g., a missing triangle and missing triad, which would be computationally demanding. This also shows how the results we obtain may strongly depend on how we modelled the data and on implementation details such as the policy to handle null values.

Correlations between different measures appear more evidently in node-aligned networks. This effect is more evident for genetic networks and co-authorship networks. In these cases, the zeros added by the alignment reinforce the correlation among the measures.


\section{Guidelines}\label{guidelines}

From our literature study, theoretical framing and experiments it appears how layer comparison measures can be very valuable and often succeed in practice to characterize the structure of multiplex networks, but they are not always straightforward to use. Therefore, in this section, we list a set of guidelines motivated by our experience acquired while testing these measures and by the results presented in the previous section.

One important aspect to consider when choosing which function to use is the distribution of values in the property matrix. Among the criteria that can be used to characterize layer property vectors and comparison functions, the following appear to be useful: 

\begin{itemize}
\item Sparsity -- A layer property vector is sparse if the number of 0s is much higher than the number of non-0 values. 
\item Degeneracy -- A layer property vector degenerates if its values are (almost) constant. Sparsity is a special case of degeneracy.
\item Linearity -- A layer property vector is linear if the values in the vector and their rank are linearly correlated.
\item Scale invariance -- a similarity function is scale invariant if it does not (significantly) change when one or more layer property vectors are multiplied by a constant.
\end{itemize}

We now list our guidelines, divided into four main areas.

\subsection{Number of measures}

The number of available measures is very large, considering that the fifty options used in our experiments are only some of the measures we can obtain using different combinations of \textit{property matrices} and observation functions. While the choice of the measures to be used for a specific empirical network is of course influenced by what the analyst is interested in, e.g., degree-based similarity, betweenness-based, or specific motifs that are motivated by the application context, our experiments show that different measures highlight different types of similarities.

The choice of which measures to use can be simplified using the correlation plots in Figures {\ref{fig:genetic_box corr}\,-\,\ref{fig:collaboration_box corr}}. Groups of measures producing highly correlated values can be identified, and one measure for each group can be chosen. In particular, JS, KL and D divergences are similar, and JS divergence can be used from this group. Jaccard, coverage and Kulczy\'nski are similar, and Jaccard or coverage can be used --- with the latter highlighting how the non-overlapping structures are distributed across the two layers, e.g., if one layer is containing the other.

When comparing layers by comparing a single value, particular attention should be paid to the so called discriminative power or uniqueness of the measure, i.e., the capability of a measure of taking different values on non-isomorphic networks \cite{konsta1996}. For example, while mean is not a representative measure for non-regular distributions, it can still be used to compare two distributions, such as degree distributions. But not alone, because the same degree does not imply the same topology.

While min can be useful in general to characterize a distribution if used together with other statistical summaries, it does not appear to be very useful to compare layers where there is typically at least one node having value 0 --- for example, min degree is 0 for all layers for most networks. On the contrary, max can be useful, e.g., to include the size of the layers in the comparison.

\subsection{Node-alignment}

The choice of whether a node-aligned or generalized multiplex model should be used is often clear from the context. For example, we would typically not align nodes when layers represent different social network sites, to represent the fact that users may not have accounts on some sites, while we would typically align nodes in a multirelational network about people interacting in multiple ways, where not having edges on a layer does not imply that the person cannot interact in that specific way.

However, the choice may have a significant influence on the results of our analysis as highlighted by our experiments.

Node-alignment may lead to some degeneracy. As expected, node-existence measures are useless, but also other cases are affected, such as measures 11-16 (degree) and 27-32 (clustering coefficient).

Measures based on node existence indicate the effect of node alignment on the other measures. So, before using link-based measures (such as edge Jaccard), it is important to check e.g., node Jaccard to understand whether comparing higher order structures is meaningful, or whether the results will just be a consequence of the amount of node overlapping across layers.

Rank correlation can suffer from node alignment because of false tie resolution, and also Pearson correlation results may become less evident, as shown by the experiments where positive and/or negative correlations are lost or decreased depending on the type of networks.

\subsection{Sparsity}

SMC and Hamann are only useful for non-sparse, non-degenerated cases, which in our experiments correspond to node existence on generalized networks. Russel-Rao also suffers if property vectors are sparse. 
As an example, these measures do not work well for triangle-existence property matrices in general.

\subsection{Linearity}

Having non-linear distributions of values in the property vectors, as it is the case for degree property matrices, is not problematic when computing linear correlation. Linear correlation (Pearson) is however often preferable to rank correlation, which can be problematic in case of generalized networks (because of null values) and also for node-aligned networks --- because of the many nodes with the same values.				

\section{Conclusion}\label{conclusions}

A summary of our guidelines is that there are many ways to compare layers, but (1) not all methods are always appropriate, and (2) some are often correlated, which means that if we only want a small number of layer similarities we can give priority to one for each group of related measures.

As we mentioned in the introduction, our framework captures several measures appeared in the literature: node activity overlapping \cite{Nicosia2015}, global overlapping of edges \cite{Bianconi2013} and absolute binary multiplexity \cite{gemmetto2015multiplexity} are applications of the Russel-Rao function to node and edge existence property vectors, average edge overlap \cite{Diakonova2015} and \cite{battiston2014structural} are respectively the Jaccard and coverage functions applied to edge existence. A general recommendation is to use the original names: all the measures used in this work and mentioned in this paragraph are applications of existing proximity measures, most of them well known to data analysts. Calling them by their name, such as edge Jaccard, makes it simpler to understand when it is reasonable to apply them if we already know the original measure.


Also, notice that our framework allows the definition of a large number of other functions not tested in this article, also considering directed/undirected networks, weights, and other mesostructures such as motifs. Other network summary functions that are not specific for multiplex networks can also be obtained as combinations of property matrices and observational functions. Examples are order (node existence + sum), size (edge existence + sum), density (edge existence + mean), average path length (dyad distance + mean), etc. We believe that splitting the problem of computing layer similarities into the two problems of (1) deciding what to observe and (2) deciding how to compare these observations using existing generic comparison functions gives the analyst the ability to easily generate custom layer comparisons that are appropriate for the problem at hand.

\section*{Data accessibility}
The experiments have been performed using the multinet library available at \url{https://cran.r-project.org/package=multinet} and twenty-three multilayer networks available at \url{http://deim.urv.cat/~manlio.dedomenico/data.php}. 

\section*{Competing interests}
We declare we have no competing interests.

\section*{Authors' contributions}
PB, AC and MM created an initial concept for the paper; All authors developed the concept to its current state and designed the experiments; PB gathered and prepared all datasets; MM implemented and executed all experiments and simulations; All authors analysed data and discussed results, drafted, critically reviewed the manuscript and approved the final version.

\section*{Funding statement}

This work was partially supported by National Science Centre Poland, the decision no. DEC-2016/21/D/ST6/02408; the European Union`s Horizon 2020 research and innovation programme under grant agreement No. 691152 (RENOIR) and 727040 (Virt-EU); and the Polish Ministry of Science and Higher Education fund for supporting internationally co-financed projects in 2016-2019 (agreement no. 3628/H2020/2016/2).



\begin{thebibliography}{9}

\bibitem{bassett2008hierarchical}
Bassett DS, Bullmore E, Verchinski BA, Mattay VS, Weinberger DR,  Meyer-Lindenberg A. 2008 Hierarchical organization of human cortical networks in health and schizophrenia. \textit{Journal of Neuroscience}, \textbf{28(37)}, 9239-9248.

\bibitem{Batagelj1995}
Batagelj V, Bren M. 1995 Comparing resemblance measures. \textit{Journal of classification}, \textbf{12(1)}, 73-90.

\bibitem{battiston2014structural}
Battiston F, Nicosia V, Latora V. 2014 Structural measures for multiplex networks. \textit{Phys. Rev. E}, \textbf{89(3)}, 032804.

\bibitem{bazzoli1999taxonomy}
Bazzoli GJ, Shortell SM, Dubbs N, Chan C, Kralovec P. 1999 A taxonomy of health networks and systems: bringing order out of chaos. \textit{Health services research}, \textbf{33(6)}, 1683.

\bibitem{Berlingerio2012}
Berlingerio M, Coscia M, Giannotti F, Monreale A, Pedreschi D. 2013 Multidimensional networks: foundations of structural analysis. \textit{World Wide Web}, \textbf{16(5-6)}, 567-593.

\bibitem{Berlingerio2013}
Berlingerio M, Pinelli F, Calabrese F. 2013 Abacus: frequent pattern mining-based community discovery in multidimensional networks. \textit{Data Mining and Knowledge Discovery}, \textbf{27(3)}, 294-320.

\bibitem{Jalili2017}
Jalili M, Orouskhani Y,Asgari M, Alipourfard N, Perc M. 2017 Link prediction in multiplex online social networks,\textit{R. Soc. open sci. \textbf{4}}: 160863.

\bibitem{Bianconi2013}
Bianconi G. 2013 Statistical mechanics of multiplex networks: Entropy and overlap. \textit{Phys. Rev. E}, \textbf{87(6)}, 062806.

\bibitem{camarinha2012taxonomy}
Camarinha-Matos LM, Afsarmanesh H. 2012 \textit{Taxonomy of Collaborative Networks Forms}. GloNet project, Draft Working Document.

\bibitem{cardillo2013emergence}
Cardillo A,  G{\'o}mez-Gardenes J, Zanin M, Romance M, Papo D, Del Pozo F, Boccaletti S. 2013 Emergence of network features from multiplexity. \textit{Sci. Rep.}, \textbf{3}. 1344 , doi:10.1038/srep01344
 
\bibitem{coleman1957diffusion}
Coleman J, Katz E, Menzel H. 1957. The diffusion of an innovation among physicians. \textit{Sociometry}, \textbf{20(4)}, 253-270.

\bibitem{Crooks2008}
Crooks, GE. 2008. Inequalities between the Jenson-Shannon and Jeffreys divergences.  \textit{Tech. Note 004}, http://threeplusone.com/pubs/technote/CrooksTechNote004.pdf

\bibitem{de2015identifying}
De Domenico M, Lancichinetti A, Arenas A, Rosvall M. 2015. Identifying modular flows on multilayer networks reveals highly overlapping organization in interconnected systems. \textit{Phys. Rev. X}, \textbf{5(1)}, 011027.

\bibitem{DeDomenico2015a}
De Domenico M, Nicosia V, Arenas A, Latora V. 2015. Structural reducibility of multilayer networks. \textit{Nature communications}, \textbf{6}, 6864.

\bibitem{de2014navigability}
De Domenico M, Sol{\'e}-Ribalta A,  G{\'o}mez S, Arenas A. 2014. Navigability of interconnected networks under random failures. \textit{Proc. Natl Acad. Sc. USA}, \textbf{111(23)}, 8351-8356.

\bibitem{Diakonova2015}
Diakonova M, Nicosia V, Latora V, San Miguel M. 2016. Irreducibility of multilayer network dynamics: the case of the voter model. \textit{New J. Phys.}, \textbf{18(2)}, 023010.

\bibitem{Faust2006ComparingStructure}
Faust K. 2006. Comparing social networks: size, density, and local structure. \textit{Metodoloski zvezki}, \textbf{3(2)}, 185.

\bibitem{gemmetto2015multiplexity}
Gemmetto V, Garlaschelli D. 2015. Multiplexity versus correlation: the role of local constraints in real multiplexes. \textit{Sci. Rep.}, \textbf{5}.

\bibitem{harland2001taxonomy}
Harland CM, Lamming RC, Zheng J, Johnsen TE. 2001. A taxonomy of supply networks. \textit{Journal of Supply Chain Management}, \textbf{37(3)}, 21-27.

\bibitem{Iacovacci2015}
Iacovacci J, Wu Z, Bianconi G. 2015. Mesoscopic structures reveal the network between the layers of multiplex data sets. \textit{Phys. Rev. E}, \textbf{92(4)}, 042806.

\bibitem{kim13}
Kim, J. Y., Goh, K.-I. 2013. Coevolution and Correlated Multiplexity in Multiplex Networks. \textit{Phys. Rev. Letters}, \textbf{111(5)}, 58702.

\bibitem{kapferer1972strategy}
Kapferer B. 1972. \textit{Strategy and transaction in an African factory: African workers and Indian management in a Zambian town}. Manchester University Press.

\bibitem{Kivela2014}
Kivel{\"a} M, Arenas A, Barthelemy M, Gleeson JP, Moreno Y, Porter MA.2014 Multilayer networks. \textit{Journal of complex networks}, \textbf{2(3)}, 203-271.

\bibitem{konsta1996}
Konstantinova EV. 1996. The Discrimination Ability of Some Topological and Information Distance Indices for Graphs of Unbranched Hexagonal Systems. \textit{Journal of Chemical Information and Computer Sciences}, \textbf{36}, 54-57.

\bibitem{krackhardt1987cognitive}
Krackhardt D. 1987. Cognitive social structures. \textit{Social networks}, \textbf{9(2)}, 109-134.

\bibitem{kullback1951information}
Kullback S, Leibler RA. 1951. On information and sufficiency. \textit{The annals of mathematical statistics}, 22(1), 79-86.

\bibitem{DBLP:conf/asonam/MagnaniR11}
Magnani M, Rossi L. 2011. The ml-model for multi-layer social networks. \textit{2011 International Conference on Advances in Social Networks Analysis and Mining (ASONAM)}, 5-12. IEEE.

\bibitem{MagnaniSBP2013b}
Magnani M, Rossi L. 2013. Formation of Multiple Networks. \textit{6th International Conference, SBP 2013}, \textbf{13}, 257-264.

\bibitem{Nicosia2013}
Nicosia V, Bianconi G, Latora V, Barthelemy M. 2013. Growing multiplex networks. \textit{.Phys. Rev. Lett}, \textbf{111(5)}, 058701.

\bibitem{Nicosia2015}
Nicosia V, Latora V. 2015. Measuring and modeling correlations in multiplex networks. \textit{Phys. Rev. E}, \textbf{92(3)}, 032805.

\bibitem{onnela2012taxonomies}
Onnela J.P, Fenn D.J, Reid S, Porter M.A, Mucha P.J, Fricker M.D, Jones NS 2012. Taxonomies of networks from community structure. \textit{Phys. Rev. E}, \textbf{86(3)}, 036104.

\bibitem{padgett1993robust}
Action R. 1993. the Rise of the Medici. \textit{American Journal of Sociology}, \textbf{98}, 1259-1319.

\bibitem{pathan2007taxonomy}
Pathan AMK, Buyya R. 2007. \textit{A taxonomy and survey of content delivery networks}. Grid Computing and Distributed Systems Laboratory, University of Melbourne, Technical Report.

\bibitem{rossi2015towards}
Rossi L, Magnani M. 2015. Towards effective visual analytics on multiplex and multilayer networks. \textit{Chaos, Solitons \& Fractals}, \textbf{72}, 68-76.

\bibitem{Salehi2015survey}
Salehi M, Sharma R, Marzolla M, Magnani M, Siyari P, Montesi D. 2015. Spreading processes in multilayer networks. \textit{IEEE Transactions on Network Science and Engineering}, \textbf{2(2)}, 65-83.

\bibitem{segaran2007programming}
Segaran T. 2007. \textit{Programming collective intelligence: building smart web 2.0 applications}. O'Reilly Media, Inc.".

\bibitem{shannon2002mathematical}
Shannon CE, Weaver W. 1998. \textit{The mathematical theory of communication}. University of Illinois press.

\bibitem{Smith2016EmpiricalSize}
Smith A, Calder CA, Browning CR. 2016. Empirical reference distributions for networks of different size. \textit{Social networks}, \textbf{47}, 24-37.

\bibitem{snijders2006new}
Snijders TA, Pattison PE, Robins GL, Handcock MS. 2006. New specifications for exponential random graph models. \textit{Sociological methodology}, \textbf{36(1)}, 99-153.

\bibitem{Soundarajan2014}
Soundarajan S, Eliassi-Rad T, Gallagher B. 2014. A guide to selecting a network similarity method. \textit{In Proceedings of the 2014 SIAM International Conference on Data Mining}, 1037-1045. Society for Industrial and Applied Mathematics.

\bibitem{stark2006biogrid}
Stark C, Breitkreutz BJ, Reguly T, Boucher L, Breitkreutz A, Tyers M. 2006. BioGRID: a general repository for interaction datasets. \textit{Nucleic acids research}, \textbf{34}, D535-D539.

\bibitem{vanWijk2010ComparingTheory}
Van Wijk BC, Stam CJ, Daffertshofer A. 2010. Comparing brain networks of different size and connectivity density using graph theory. \textit{PloS one}, \textbf{5(10)}, e13701.

\bibitem{vickers1981representing}
Vickers M, Chan S. 1981. \textit{Representing classroom social structure}. Victoria Institute of Secondary Education, Melbourne.

\bibitem{wang2012statistical}
Wang T, Krim H. 2012. Statistical classification of social networks. \textit{In 2012 IEEE International Conference on Acoustics, Speech and Signal Processing (ICASSP)} 3977-3980. IEEE.

\bibitem{wiener1947structural}
Wiener H. 1947. Structural determination of paraffin boiling points. \textit{Journal of the American Chemical Society}, \textbf{69(1)}, 17-20.





\end{thebibliography}
\end{document}